\documentclass[amssymb,amsmath,preprint,superscriptaddress]{revtex4-1}
\usepackage[colorlinks]{hyperref}
\usepackage{graphicx}
\usepackage{multirow}

\usepackage{slashed}
\usepackage{gensymb}
\usepackage[compat=1.1.0]{tikz-feynman}
\usepackage{orcidlink}
\begin{document}
\title{Searching for Axial Neutral Current Non-Standard Interactions of neutrinos by DUNE-like experiments}

\author{S. Abbaslu\, \orcidlink{0000-0003-3567-717X}}
\email{s-abbaslu@ipm.ir}
\affiliation{School of Physics, Institute for Research in Fundamental Sciences (IPM), PO Box 19395-5531, Tehran, Iran}

\author{M. Dehpour\, \orcidlink{0000-0001-9706-9984}}
\email{m.dehpour@mail.sbu.ac.ir}
\affiliation{Department of Physics, Shahid Beheshti University,
PO Box 19839-63113, Tehran, Iran}

\author{Y. Farzan\, \orcidlink{0000-0003-0246-5349}}
\email{yasaman@theory.ipm.ac.ir}
\affiliation{School of Physics, Institute for Research in Fundamental Sciences (IPM), PO Box 19395-5531, Tehran, Iran}

\author{S. Safari\, \orcidlink{0000-0002-9196-3464}}
\email{sa.safari@mail.sbu.ac.ir}
\affiliation{Department of Physics, Shahid Beheshti University,
PO Box 19839-63113, Tehran, Iran}

\begin{abstract}
    The increasingly precise neutrino experiments raise the hope for searching for new physics through studying the impact of Neutral Current (NC) Non-Standard Interactions (NSI) of neutrinos with matter fields. Neutrino oscillation experiments along with the Elastic Coherent $\nu$ Nucleus Scattering (CE$\nu$NS) experiments already set strong bounds on all the flavor elements of the ``vector" NC NSI. However, ``axial" NC NSI can hide from these experiments. We show how a DUNE-like experiment can probe these couplings by studying NC Deep Inelastic Scattering (DIS) events. We find that strong bounds can be set on the axial NC NSI of neutrinos with the $u$, $d$, and $s$ quarks. We show that using both the near and far detectors, a DUNE-like experiment can significantly improve the present bounds on all the flavor elements. 

\begin{center}
\textit{This work is dedicated to the memory of our dear colleague Thomas Weiler who passed away on the day of submission of this article to the arXiv.  }
\end{center}
\end{abstract}
\maketitle

\section{Introduction}
\label{sec:intro}
Searching for Non-Standard neutrino Interactions (NSI) opens a door towards new physics. In recent years, with the prospect of precision experiments such as DUNE, there has been a renewed interest in the neutral current NSI both from an experimental and from a model building point of view (for an incomplete list of references, see \cite{Farzan:2017xzy,Escrihuela:2009up,Bernal:2022qba, deSalas:2021aeh, Terol-Calvo:2019vck, Farzan:2019xor, Denton:2018dqq, Bakhti:2016prn, Farzan:2015hkd, Farzan:2015doa, Bakhti:2014pva,Flores:2018kwk, Dutta:2017nht, Miranda:2015dra,Salvado:2016uqu,Bischer:2018zcz,Farzan:2016wym,Masud:2021ves,Masud:2021epj,Proceedings:2019qno,Masud:2018pig,Masud:2016nuj,Chatterjee:2014gxa,Esmaili:2013fva,Formozov:2019oah,Weatherly:2017ojv,Khan:2017oxw,Deepthi:2016erc,Ge:2016dlx,Stapleford:2016jgz,IceCube:2022pbe,Coloma:2023ixt,Amaral:2023tbs,Karkkainen:2023ozw, Denton:2022nol,Coloma:2022umy, Arguelles:2022tki,Coloma:2022avw, Coloma:2020gfv, Babu:2020nna,Giunti:2019xpr, Arguelles:2019xgp,Coloma:2015kiu, Coloma:2016gei,Coloma:2017egw,Coloma:2017ncl, AristizabalSierra:2017joc,Denton:2018xmq, Chaves:2018sih,AristizabalSierra:2019zmy, Chatterjee:2022nia,Capozzi:2020jqb, Chatterjee:2020kkm, Capozzi:2019iqn, Agarwalla:2016fkh, Palazzo:2011vg, DeRomeri:2022twg, Delgadillo:2023lyp, Chatterjee:2022mmu,KM3NeT:2021nnf,Cherchiglia:2023aqp,Eramzhian:1974uu,Folomeshkin:1976vj}). The Neutral Current (NC) NSI can be parameterized as
\begin{align}
    \frac{G_{\rm F}}{\sqrt{2}} \left[\overline{\nu}_\alpha \gamma^\mu \left(1-\gamma_5\right) \nu_\beta\right]\left[\overline{f} \gamma_\mu \left(\epsilon_{\alpha \beta}^{Vf}+\epsilon_{\alpha \beta}^{Af}\gamma_5\right)f\right] \quad {\rm where} \quad f\in\{ e , u, d , s\},
\end{align}
where $\epsilon^{Vf}$ and $\epsilon^{Af}$ are dimensionless $3\times 3$ symmetric matrices in the flavor space, respectively, determining the vector and axial NSI. $G_F$ is the Fermi constant.
The $\epsilon^{Vf}$ couplings have been extensively studied in the literature because the neutrino propagation in matter as well as Coherent Elastic neutrino Nucleus Scattering (CE$\nu$NS) are sensitive to $\epsilon^{Vf}$ \cite{Esteban:2018ppq,Barranco:2005yy}. However, obtaining information on the axial NSI is more challenging, as the $\epsilon^{Af}$ couplings do not affect the neutrino oscillation patterns or CE$\nu$NS.
The Deuterium dissociation with a neutrino beam ($D+\nu \to n+p+\nu$), being a Gamow-Teller process, is sensitive to a combination of the $\epsilon^{Af}$ couplings. The SNO data on the Neutral Current (NC) measurement of the solar neutrino flux therefore yields information on a certain combination of $\epsilon^{Af}$. The NuTeV neutrino nucleus scattering experiment \cite{NuTeV:2001whx} has set strong bounds on the $\epsilon^{Au}_{\mu \alpha}$ and $\epsilon^{Ad}_{\mu \alpha}$ elements. There are also bounds from the CHARM experiment \cite{CHARM:1986vuz,CHARM-II:1994dzw} on the $\epsilon^{Au}_{e \alpha}$ and $\epsilon^{Ad}_{e \alpha}$ elements.
However, as we shall discuss in sect.~\ref{sec:Bounds}, large values of $\epsilon^{Af}_{\tau\tau}\sim \mathcal{O}(1)$ are still experimentally allowed. $\epsilon^{Af}$ can also be probed via the scattering of higher energy beams at detectors. Ref. \cite{Escrihuela:2023sfb} has studied the bounds to be set by FASER$\nu$ on $\epsilon^{Af}_{\mu \alpha}$. 

In this paper, we shall discuss how NC events in a DUNE-like experiment can provide information on $\epsilon^{Af}$.
The Near Detector (ND) of an experiment such as DUNE will receive a huge flux of $\nu_\mu$ or $\overline{\nu}_\mu$ making it possible to probe tiny $\epsilon_{\mu \alpha}^{Aq}$ values. At the Far Detector (FD) of such an experiment, the oscillated flux of neutrinos will also contain $\nu_e$ (or $\overline{\nu}_e$) and $\nu_\tau$ (or $\overline{\nu}_\tau$) components, making it possible to probe the $\epsilon_{ee}^{Aq}$, $\epsilon_{e\tau}^{Aq}$ and $\epsilon_{\tau \tau}^{Aq}$ elements. In particular, as we shall see, the nearly unconstrained $\epsilon_{\tau \tau}^{Aq}$ element can be strongly probed by the far detector of a DUNE-like experiment. For the first time, we discuss axial NC NSI with the $s$ quark and forecast the bounds that an experiment such as  DUNE can set on 
$\epsilon_{\alpha \beta}^{As}$.

The present paper is organized as follows. In sect.~\ref{sec:Bounds}, we summarize the present bounds on $\epsilon_{\alpha \beta}^{Aq}$. In Sect.~\ref{sec:scattering},  we derive the cross sections of NC Deep Inelastic Scattering (DIS) of neutrinos off nucleons both within the Standard Model (SM) and in the presence of NSI. We also discuss the impact of NSI on the NC resonance scattering.  In sect.~\ref{sec:NCNSI}, we compute the number of NC DIS events in the neutrino and antineutrino modes at both ND and FD and discuss the sensitivity to the NSI parameters. In sect.~\ref{sec:forecast}, we forecast the bounds by a DUNE-like experiment on $\epsilon_{\alpha \beta}^{Aq}$. Results are summarized in sect.~\ref{sec:summary}.

\section{Bounds from  past scattering experiments}
\label{sec:Bounds}
In subsect.~\ref{CHARM-NuTeV}, we enumerate the bounds from NuTeV and CHARM scattering experiments on $\epsilon_{\alpha \beta}^{qA}$ as well as the constraints from the loop effects.  In subsect.~\ref{deutrerium}, we review the bounds from the SNO experiment on the axial NSI.

\subsection{Bounds from CHARM and NuTeV scattering experiments}
\label{CHARM-NuTeV}
Ref.~\cite{Davidson:2003ha} has recast the NC and CC neutrino nucleus scattering measurement results by the neutrino scattering experiments, CHARM and NuTeV as bounds on the NSI couplings (see also Ref.~\cite{Escrihuela:2009up}):
\begin{align} 
    {\rm From \ NuTeV:} \quad |\epsilon_{\mu \mu}^{Au}|<0.006, \quad |\epsilon_{\mu \mu}^{Ad}|<0.018, \quad |\epsilon_{\mu \tau}^{Au}|,|\epsilon_{\mu \tau}^{Ad}|<0.01,
    \label{eq:nuTEV}
\end{align}
and
\begin{align} 
    {\rm From \ CHARM:} \quad |\epsilon_{e e}^{Au}|<1, \quad |\epsilon_{e e}^{Ad}|<0.9, \quad |\epsilon_{e \tau}^{Au}|,|\epsilon_{e \tau}^{Ad}|<0.5.
    \label{CHARM} 
\end{align}
Two comments are in order:
\begin{itemize}
\item These bounds apply only if the mass of the mediator responsible for NSI is much heavier than few GeV. Otherwise, relying on the four-Fermi effective interactions is not justified.
\item The bounds on the $\mu\mu$ and $\mu \tau$ elements are very stringent. To avoid uncertainties in the neutrino fluxes and Parton Distribution Functions (PDFs), Ref.~\cite{Davidson:2003ha} uses the ratio of Neutral Current NC to Charged Current (CC) events which, in the first approximation for an isospin singlet target, is independent of PDFs.
Remember that while the cross section of NC interaction of $\nu$ or $\overline{\nu}$ off the $s$- or $\overline{s}$-quark is equal to that for the scattering off the $d$- or $\overline{d}$-quark ({\it i.e.,} $\sigma_{\rm NC}(\nu+s\to \nu+s)/\sigma_{\rm NC}(\nu+d\to \nu+d)\simeq 1$), the same ratio for the CC interactions is suppressed by the sine square of the Cabbibo angle:
$\sigma_{\rm CC}(\nu+s\to \nu+s)/\sigma_{\rm CC}(\nu+d\to \nu+d)\simeq \sin^2 \theta_{\rm C}$. As a result, the approximation used to derive the bounds in Eq.~(\ref{eq:nuTEV}) is broken by the presence of the $s$ and $\overline{s}$ sea quarks by $\int dx\, x \left[f_s(x)+f_{\overline{s}}(x)\right]/\int dx\, x \left[f_d(x)+f_{\overline{d}}(x)\right]$ which, according to table \ref{tab:moments}, is about $\mathcal{O}(15 \%)$
\end{itemize}
At the one-loop level,  off-diagonal elements of NSI coupling can lead to the charged lepton flavor conversion on nuclei. The bound on ${\rm Ti}+\mu \to {\rm Ti} +e$ implies the stringent constraint \cite{Davidson:2003ha}:
\begin{align} 
    |\epsilon_{e \mu }^{qR}|,|\epsilon_{e \mu }^{qL}|<7.7 \times 10^{-4} \ ,\ \ \ \ q\in \{ u,d \}.
\end{align}
The weakest bound is on the $\tau \tau$ element, which again comes from the loop induced invisible decay of the $Z$ boson \cite{Davidson:2003ha}:
\begin{align} 
    |\epsilon_{\tau \tau}^{qA}|<\mathcal{O}(1) \ , \ \ \ \ q\in \{ u,d \}.
\end{align}

\subsection{Deuterium dissociation}
\label{deutrerium}
The $\epsilon^{Au}$ and $\epsilon^{Ad}$ couplings of the quarks induce interaction of  form
\begin{align}
    \frac{G_{\rm F}}{\sqrt{2}} \left[\overline{\nu}_\alpha \gamma^\mu \left(1-\gamma_5\right) \nu_\beta\right]\left(\epsilon_{\alpha \beta}^{An}~\overline{n}  \gamma^\mu\gamma_5 n+\epsilon_{\alpha \beta}^{Ap}~\overline{p}\gamma^\mu\gamma_5p\right),
\end{align}
where 
\begin{align}
    \epsilon_{\alpha \beta}^{Ap}&= \Delta_u \epsilon_{\alpha \beta}^{Au}+\Delta_d \epsilon_{\alpha \beta}^{Ad}, \notag \\
    \epsilon_{\alpha \beta}^{An}&= \Delta_u \epsilon_{\alpha \beta}^{Ad}+\Delta_d \epsilon_{\alpha \beta}^{Au}.
\end{align}
The values of $\Delta_q$ can be found in table 4 of Ref. \cite{Cirelli:2013ufw}. For reference, we take $\Delta_u=0.84$ and $\Delta_d=-0.43$ \cite{Belanger:2013oya}. The Deuterium is an isospin singlet. As a result, its dissociation is sensitive only to the combination 
\begin{align}
    \epsilon^{An}_{\alpha \beta}-\epsilon^{Ap}_{\alpha \beta}=\left(\Delta_u-\Delta_d\right)\left(\epsilon^{Au}_{\alpha \beta}-\epsilon^{Ad}_{\alpha \beta}\right)=1.27 \left(\epsilon^{Au}_{\alpha \beta}-\epsilon^{Ad}_{\alpha \beta}\right).
\end{align}
In table 4 of Ref. \cite{Coloma:2023ixt}, bounds from SNO on $\epsilon^{Au}_{\alpha \beta}-\epsilon^{Ad}_{\alpha \beta}$ can be found. In fact, Ref.~\cite{Coloma:2023ixt} has found solutions for SNO in the presence of $\epsilon^{Af}$. Of course, for each $\epsilon^{Au}_{\alpha \beta}-\epsilon_{\alpha \beta}^{Ad}$, there is a solution including zero, with an upper bound of $\sim 0.1$ but, due to the interference with the SM amplitudes, there are also non-trivial solutions deviating from zero: {\it e.g.,} $-2.1<\epsilon^{Au}_{ee}-\epsilon^{Ad}_{ee}<-1.8$ or $1.6 <\epsilon^{Au}_{\mu \tau}-\epsilon^{Ad}_{\mu \tau}<1.9$. At first sight, we do not expect interference for off-diagonal elements as the SM is lepton flavor conserving. However, we should remember that solar neutrinos interacting at SNO are incoherent mass eigenstates rather than flavor eigenstates; hence, the interference takes place also for off-diagonal $\epsilon^{Af}$.
Under the assumption that the mediator of NSI is heavier than few GeV, the bounds from SNO are not competitive with the scattering bounds enumerated in subsect.~\ref{CHARM-NuTeV}. The only non-trivial SNO solution that survives the CHARM and NuTeV bounds is
\begin{align}
    -1.6<\epsilon^{Au}_{\tau \tau}- \epsilon^{Ad}_{\tau \tau}<-1.4.
    \label{eq:surviving}
\end{align}
We shall discuss how a DUNE-like experiment can test this surviving solution.

\section{Neutrino scattering in the presence of Neutral Current Axial NSI}
\label{sec:scattering}
At energies much higher than $\sim$10 GeV,  such as the energy of the neutrino beam at the LHC, cross sections can be computed using the Deep Inelastic Scattering (DIS) formulas. However, at energies of few GeV which is of relevance to a DUNE-like experiment, DIS and resonance scattering can have comparable contributions and both should be taken into account. In this section, we discuss the effects of NSI on the cross section of both types of scattering one by one.

\subsection{Deep Inelastic Scattering (DIS) in the presence of NSI}
\label{sec:DIS}
In this sub-section, we first derive the cross section of NC DIS for neutrinos and antineutrinos of a definite flavor in the presence of NSI. We then discuss how we can compute the NC DIS cross sections for a coherent linear combination of flavors such as the neutrino states arriving at the far detector after undergoing oscillation.

Combining the standard and non-standard interactions, we can write the  effective NC Lagrangian as follows:
\begin{align}
    \mathcal{L}_{\text{tot}}^{\rm NC} = -\frac{G_{\rm F}}{\sqrt{2}} \sum_{\alpha,\beta,q} \left[ \overline{\nu}_\alpha \gamma^\mu \left(1 - \gamma_5\right) \nu_\beta \right] \left[ \overline{q}\gamma_{\mu}\left(f^{Vq}_{\alpha\beta} + f^{Aq}_{\alpha\beta} \gamma_5 \right)q\right],
\end{align}
where the modified vector and axial-vector parameters  $f^{Vq}_{\alpha\beta}$ and $f^{Aq}_{\alpha\beta}$ for the quarks are given by
\begin{align}
    f^{Vq}_{\alpha\beta} = \epsilon^{Vq}_{\alpha\beta} + g^{Vq}\delta_{\alpha\beta} \qquad \text{and} \qquad f^{Aq}_{\alpha\beta}= \epsilon^{Aq}_{\alpha\beta} + g^{Aq}\delta_{\alpha\beta}.		
\end{align}
The values of the Standard Model (SM) coefficients $g^{L,R}$ for the SM fermions as well as $g^{A,V}$ are given in table \ref{gg}.
Remember that $\epsilon^{V/A q}_{\alpha\beta}= \epsilon^{Lq}_{\alpha\beta} \pm \epsilon^{Rq}_{\alpha\beta}$  and $g^{V/A q}_{\alpha\beta} = g^{Lq} \pm g^{Rq}$.  

\begin{table}[h]
    \caption{The standard Neutral Current (NC) coefficients for the quarks and leptons. Here, $\theta_{\rm W}$ denotes the Weinberg angle. \label{gg}}
    \begin{ruledtabular}
        \begin{tabular}{c c c c c}
            & Up type quarks & Down type quarks & Charged leptons & Neutral leptons \\
            & ($u,c,t$) & ($d,s,b$) & ($e,\mu,\tau$) & ($\nu_e,\nu_\mu,\nu_\tau$) \\
            \colrule
            $g^L$ & $\frac{1}{2}-\frac{2}{3}\sin^2\theta_{\rm W}$ & $-\frac{1}{2}+\frac{1}{3}\sin^2\theta_{\rm W}$ & $-\frac{1}{2}+\sin^2\theta_{\rm W}$ & $\frac{1}{2}$\\
            $g^R$ & $-\frac{2}{3}\sin^2\theta_{\rm W}$ & $\frac{1}{3}\sin^2\theta_{\rm W}$ & $\sin^2\theta_{\rm W}$ & $0$\\
            $g^V$ & $\frac{1}{2}+\frac{4}{3}\sin^2\theta_{\rm W}$ & $-\frac{1}{2}+\frac{2}{3}\sin^2\theta_{\rm W}$ & $-\frac{1}{2}+2\sin^2\theta_{\rm W}$ & $\frac{1}{2}$\\
            $g^A$ & $\frac{1}{2} $ & $-\frac{1}{2}$ & $-\frac{1}{2} $ & $\frac{1}{2}$ \\
        \end{tabular}
    \end{ruledtabular}
\end{table}

Let us now discuss the cross section of the Deep Inelastic Scattering (DIS) of neutrinos and antineutrinos of flavor $\alpha$ off nucleons as shown in Fig.~\ref{fig:kinematics1}:

\begin{figure}[b]
    \centering
    \begin{tikzpicture}
        \begin{feynman}
            \fill[black!100] (0,0) ellipse (0.13cm and 0.32cm);

            \vertex (a);
            \vertex [above right = 2cm of a] (b) {\(\overset{(-)}{\nu_{\beta}}\)};
            \vertex [below right = 2cm of a] (c) {\(\)};
            \vertex [above left = 2cm of a] (d) {\(\overset{(-)}{\nu_{\alpha}}\)};
            \vertex [below left = 2cm of a] (e) {\(\)};
            \vertex [below = 0.2cm of e] (j) {\(\)};
            \vertex [below = 0.2cm of c] (k) {\(\)};
            \vertex [below = 0.2cm of j] (l) {\(\)};
            \vertex [below = 0.2cm of k] (m) {\(\)};

            \diagram*{
                (d) -- [fermion, momentum=\(p_1\)] (a) -- [fermion] (c),
                (e) -- [fermion, momentum=\(x p_2\)] (a) -- [fermion, momentum'=\(p_3\)] (b),
                (j) -- [fermion, bend left] (k),
                (l) -- [fermion, bend left] (m),
            };
            
            \draw [decoration={brace}, decorate] (l.south west) -- (e.north west) node [pos=0.5, left] {\(N\)};
            \draw [decoration={brace}, decorate] (c.north east) -- (m.south east) node [pos=0.5, right] {\(X\)};
        \end{feynman}
    \end{tikzpicture}
    \caption{Kinematics in non-standard NC neutrino or antineutrino DIS.}
    \label{fig:kinematics1}
\end{figure}
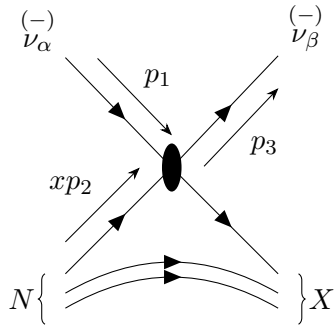

\begin{align}
    \nu_\alpha (p_1)  + N(p_2) \rightarrow \nu_\beta (p_3) + X(p') \qquad \text{where} \qquad N = n, p,
    \label{reactionNU}
\end{align}
or
\begin{align}
    \overline{\nu}_\alpha (p_1) + N(p_2) \rightarrow  \overline{\nu}_\beta(p_3) + X(p') \qquad \text{where} \qquad N = n, p,
    \label{reactionANTINU}
\end{align}
where $p_1$ and $p_3$ represent the four-momenta of the incoming and outgoing (anti)neutrinos and $p_2$ represents the four-momentum of the incoming nucleon. $p'$ is the four-momentum of the jet of hadrons produced in the final state. Inclusively, we sum  over the hadronic final states denoted by $X$. In the rest frame of the target nucleon, these four momenta are represented as follows:
\begin{align*}
    p_1^\mu &= (p_1^0, \vec{p}_1), \text{ where } |\vec{p}_1| = p_1^0 = E_{\nu}, \\
    p_3^\mu &= (p_3^0, \vec{p}_3), \text{ where } |\vec{p}_3| = p_3^0 = E^{\prime}_{\nu}, \\
    p_2^\mu &= (p_2^0, \vec{p}_2) = (M_N, 0, 0, 0).
\end{align*}
where $M_N$ represents the nucleon mass. Following the general convention in DIS studies, we show the fraction
of the nucleon momentum carried by the initial interacting parton by $x$, specifically defined as $$x = \frac{-q^2}{2p_2 \cdot q} = \frac{Q^2}{2M_N (E_\nu-E'_\nu)},$$ where $q^{\mu} = (p_1 - p_3)^{\mu}$ denotes the four-vector of momentum transfer and $Q^2=-q^2$. The $x$ parameter is called the Bjorken variable. Again, following the conventional notation,  the variable $y$ is defined as 
\begin{align*}
    y = 1 - \frac{E_\nu^{\prime}}{E_{\nu}},
\end{align*}
where $E_\nu^{\prime}$ corresponds to the outgoing neutrino energy. The scaling variables $x$ and $y$ lie in the following range: 
\begin{align*}
    0 \leq x \leq 1  \quad {\rm and} \quad  0 \leq y \leq \frac{1}{1 + {M_N x}/(2E_{\nu})}.
\end{align*}
Although rather cumbersome, it is straightforward to show that the differential cross section in terms of the parton distribution functions, $f_N^q(x)$ and $f_N^{\overline{q}}(x)$ can be written as
\begin{align}
    &\frac{d^2\sigma_{\rm NC}(\stackrel{(-)}{\nu_\alpha}  N \to \stackrel{(-)}{\nu_\beta}+X)}{dxdy}=\sigma^{0}_{\rm NC} \Bigg\{ \frac{1}{2}\left( xy^2+2x-2xy-\frac{M_N}{E_{\nu}}x^2y\right)  \notag \\ 
    &\qquad \ \times \left[\sum_{q}f_{N}^{q}(x)\left( \left|{f}^{Vq}_{\alpha\beta} \right|^2 +\left|{f}^{Aq}_{\alpha\beta} \right|^2\right)+\sum_{\overline{q}}f_{N}^{\overline{q}}(x)\left(\left|{f}^{Vq}_{\alpha\beta} \right|^2+ \left|{f}^{Aq}_{\alpha\beta} \right|^2  \right) \right] \notag \\
    & \ \ \pm2xy\left(1-\frac{y}{2}\right) \left[ \sum_{q}f_{N}^{q}(x) \Re \left[{f}^{Vq}_{\alpha\beta} (f^{Aq}_{\alpha\beta})^* \right]-\sum_{\overline{q}}f_{N}^{\overline{q}}(x) \Re \left[ {f}^{Vq}_{\alpha\beta} (f^{Aq}_{\alpha\beta})^* \right] \right] \Bigg\},
    \label{eq-dis1new}
\end{align}
with
\begin{align}
    \sigma^{0}_{\rm NC}=\frac{G_{\rm F}^2 }{\pi}\left( M_{N} E_\nu\right).
\end{align}
The plus and minus signs in the last line of Eq.~(\ref{eq-dis1new}) are respectively for neutrino and antineutrino scattering off nucleon $N$. Notice that at a DUNE-like experiment, $Q^2\sim {\rm few~GeV}^2\ll m_Z^2$. The implicit assumption in these formulas is that the mediators of NSI are also much heavier than few GeV so we can use the four-Fermi formalism. At a DUNE-like experiment, variation of $Q^2$ is limited to below $10~$GeV$^2$ so we can safely neglect the dependence of the parton distribution functions on $Q^2$. Invoking the isospin symmetry, we can write
\begin{align}
    &f_{n}^{d}(x) = f_{p}^{u}(x) \equiv u(x), \quad f_{n}^{\overline{d}}(x) = f_{p}^{\overline{u}}(x) \equiv \overline{u}(x),\notag \\
    &f_{n}^{u}(x) = f_{p}^{d}(x) \equiv d(x), \quad f_{n}^{\overline{u}}(x) = f_{p}^{\overline{d}}(x) \equiv \overline{d}(x),\notag \\
    &f_{n}^{s}(x) = f_{p}^{s}(x) \equiv s(x), \quad f_{n}^{\overline{s}}(x) = f_{p}^{\overline{s}}(x) \equiv \overline{s}(x).\quad  
    \label{eq:distribuition-1-1}
\end{align}

To obtain the total cross section, we should integrate over $y$ and $x$. The upper limit on $y$ ({\it i.e.,} $y_{\rm max}$) depends on $x$. We may expand it as $y_{\text{max}} \approx 1 - \frac{M_{N} x}{2 E_{\nu}} + \left(\frac{M_{N} x}{2 E_{\nu}}\right)^2 + \dots$.  Neglecting the higher terms induces an error of  $\mathcal{O}(5\times 10^{-4})$ which is acceptable within the uncertainties of the present analysis. Using this approximation, we can write the cross section of the scattering of the neutrino or the antineutrino off the proton, $\sigma_p$ as 
\begin{align}
    &\sigma_{p}(\stackrel{(-)}{\nu_\alpha} +p\to \stackrel{(-)}{\nu_\beta}+X) \simeq \sigma^{0}_{\rm NC}\int_{0}^{1}dx \notag \\
    &\quad \times \Bigg\{ \frac{2}{3}\left[1-\frac{3}{2} \frac{M_{p} x}{2 E_{\nu}}+\frac{9}{4}\left(\frac{M_{p} x}{2 E_{\nu}}\right)^2 \right] x \bigg[ \left[u(x)+\overline{u}(x)\right] \left( |{f}^{Vu}_{\alpha\beta}|^2 +|{f}^{Au}_{\alpha\beta} |^2\right) \notag \\
    &  \qquad \ \  +\left[d(x)+\overline{d}(x)\right] \left( |f^{Vd}_{\alpha\beta}|^2 +|f^{Ad}_{\alpha\beta} |^2\right)+\left[s(x)+\overline{s}(x)\right] \left( |{f}^{Vs}_{\alpha\beta}|^2 +|{f}^{As}_{\alpha\beta} |^2\right) \bigg]\notag \\
    &\quad \ \ \pm \frac{2}{3}\left[1-\frac{3}{2} \frac{M_{p} x}{2 E_{\nu}} +\frac{3}{2}\left(\frac{M_{p} x}{2 E_{\nu}}\right)^2 \right] x \bigg[ \left[u(x)-\overline{u}(x)\right] \Re \left[{f}^{Vu}_{\alpha\beta}  ({f}^{Au}_{\alpha\beta})^*\right] \notag \\
    &\qquad \ \ +\left[d(x) -\overline{d}(x)\right]\Re \left[f^{Vd}_{\alpha\beta}  (f^{Ad}_{\alpha\beta})^*\right] +[s(x)-\overline{s}(x)]\Re \left[f^{Vs}_{\alpha\beta}  (f^{As}_{\alpha\beta})^*\right] \bigg] \Bigg\},
    \label{eq:sigtot}
\end{align}
where the plus (the minus) sign is for the neutrinos (antineutrinos). As discussed before, the cross section of scattering off the neutron, $\sigma_n$ is obtained with $u(x) \leftrightarrow d(x).$

In order to compute the quark distribution functions, we invoke the CT18 next-to-next-to-leading order (CT18NNLO)  \cite{Hou:2019efy}  (see also \cite{Carrazza:2014gfa,Bertone:2013vaa}). The CT18 PDF set is obtained by the CTEQ-TEA collaboration implementing a comprehensive range of high-precision Large Hadron Collider (LHC) data, plus the combined HERA I+II Deep Inelastic Scattering (DIS) data, along with   the CT14 global QCD analysis. With PDF set at a given value of momentum transfer, $Q = 2 \ \text{GeV}$, we can compute the moments of the PDF such as $\int x^n u(x) dx $. The quantities entering the cross section formulas are shown in table \ref{tab:moments}.
\begin{table}[h]
   \caption{Integral of $\int_{0}^{1} dx\, x^n\left[q(x)\pm\overline{q}(x)\right]$ at $Q = 2 , \text{GeV}$ for quarks of type $u$, $d$, and $s$ with $n=1,2,3$.  We have computed the quark distribution functions $q(x)$ and $\overline{q}(x)$  using the CT18NNLO PDF  \cite{Hou:2019efy} (see also \cite{Carrazza:2014gfa,Bertone:2013vaa}). \label{tab:moments}}
    \begin{ruledtabular}
        \begin{tabular}{ c c c c}
            Integral&$u$ & $d$ & $s$ \\
            \colrule
            $\int_{0}^{1} dx\, x\left[q(x)+\overline{q}(x)\right]$ & $0.349 \pm 0.007 $& $0.193 \pm 0.007$ & $0.033 \pm 0.008$ \\
            $\int_{0}^{1} dx\, x^2\left[q(x)+\overline{q}(x)\right]$ & $0.090 \pm 0.002$ & $0.037\pm 0.001$ & $0.002\pm 0.0008$ \\
            $\int_{0}^{1} dx\, x^3\left[q(x)+\overline{q}(x)\right]$ & $0.034\pm0.0009$ &$0.012\pm0.0007$ & $0.0005\pm0.0005$ \\
            $\int_{0}^{1} dx\, x\left[q(x)-\overline{q}(x)\right]$ & $0.290 \pm 0.008$ & $0.120 \pm 0.003$ & $0.0$ \\
            $\int_{0}^{1} dx\, x^2\left[q(x)-\overline{q}(x)\right]$ & $0.084\pm0.002$ & $0.030\pm0.001$ & $0.0$ \\
            $\int_{0}^{1} dx\, x^3\left[q(x)-\overline{q}(x)\right]$& $0.033\pm0.0009$ & $0.010\pm0.0007$& $0.0$\\
        \end{tabular}
    \end{ruledtabular}
\end{table}

We expect the total cross section of scattering of nucleus to be an incoherent summation of contributions arising from neutrino (antineutrino) scattering on both protons and neutrons inside the nucleus. Up to corrections of $\mathcal{O}(20 \%)$ from shadowing, anti-shadowing and EMC effects, this is a good approximation \cite{Altmannshofer:2018xyo}. Thus, the total cross section of the scattering off a nucleus can be expressed as the sum of the individual cross sections for scattering off the ingredient protons and neutrons of that nucleus.

The flux of neutrinos reaching the near detector is expected to be a pure $\nu_\mu$ or $\overline{\nu}_\mu$ flux.
NC NSI that we have focused on cannot affect the neutrino production (pion decay) at source so the prediction of the SM for the fluxes of neutrinos also applies to the case under our study. The above formulas for the cross sections of neutrino and antineutrinos of definite flavor can therefore readily be used to compute the NC DIS rates at the near detector. However, neutrinos oscillate on their way to the far detector and as a result, they will be a coherent combination of flavor eigenstates. As discussed before axial NSI does not affect the neutrino propagation so using the standard oscillation formulas, we can write
\begin{align}
    |\nu_{\rm far}(E_\nu)\rangle =\sum_i \sum_\beta e^{i m_{Mi}^2L/(2E_\nu)}(U_{\mu i}^M)^* U_{\beta i}^M |\nu_\beta\rangle \equiv \sum_\beta\mathcal{A}_\beta|\nu_\beta \rangle  \quad {\rm (neutrino\ mode)}
\end{align} 
and
\begin{align}
    |\overline{\nu}_{\rm far}(E_\nu)\rangle =\sum_i \sum_\beta e^{i \overline{m}_{Mi}^2L/(2E_\nu)}(\overline{U}_{\mu i}^M)^*\overline{U}_{\beta i}^M |\overline{\nu}_\beta\rangle \equiv \sum_\beta\overline{\mathcal{A}}_\beta|\overline{\nu}_\beta \rangle  \quad {\rm (antineutrino\ mode)} 
\end{align}
where $U_{\beta i}^M$ ($\overline{U}_{\beta i}^M$) and $m_{Mi}$ ($\overline{m}_{Mi}$) are respectively effective mixing and mass eigenvalues of neutrinos (antineutrinos) in matter. Notice that $|\mathcal{A}_\beta|^2=P(\nu_\mu\to \nu_\beta)$ and $|\overline{\mathcal{A}}_\beta|^2=P(\overline{\nu}_\mu\to \overline{\nu}_\beta)$ so 
\begin{align*}
    \sum_\alpha |\mathcal{A}_\alpha|^2=1 \quad {\rm and} \quad \sum_\alpha |\overline{\mathcal{A}}_\alpha|^2=1 .
\end{align*}
Thus,
\begin{align} 
    \mathcal{M}(\nu_{\rm far} +q \to \nu_\alpha +q)=& \sum_\beta \mathcal{A}_\beta \mathcal{M}(\nu_{\beta} +q \to \nu_\alpha +q), \notag \\
    \mathcal{M}(\overline{\nu}_{\rm far} +q \to \overline{\nu}_\alpha +q)=& \sum_\beta \overline{\mathcal{A}}_\beta  \mathcal{M}(\overline{\nu}_{\beta} +q \to \overline{\nu}_\alpha +q).
\end{align}
As usual, by squaring these amplitudes, the corresponding cross sections can be computed. However, it is more convenient to switch from the flavor eigenstates ($\nu_\alpha$ with $\alpha \in \{e,\mu,\tau\}$) to the eigenstates at the far detector: $\nu_\alpha$ with $\alpha \in \{{\rm far},\perp,T\}$ where $\nu_\perp$ and $\nu_T$ are two arbitrary neutrino states perpendicular to $\nu_{\rm far}$ and each other: 
\begin{align*}
    \langle \nu_\perp|\nu_{\rm far}\rangle=\langle \nu_\perp|\nu_{T}\rangle=\langle \nu_T|\nu_{\rm far}\rangle=0.
\end{align*}
Without loss of generality, we can choose
\begin{align} 
    \left[ \begin{matrix} \nu_{\rm far}\cr \nu_\perp\cr \nu_T \end{matrix}\right]=\left[\begin{matrix}
    \mathcal{A}_e &
    \mathcal{A}_\mu &
    \mathcal{A}_\tau\cr 0 & -\mathcal{A}_\tau^*/\mathcal{A} & \mathcal{A}_\mu^*/\mathcal{A} \cr \frac{
    \mathcal{A}\mathcal{A}_e}{|\mathcal{A}_e|} & -\frac{\mathcal{A}_\mu|\mathcal{A}_e|}{\mathcal{A}} & -\frac{\mathcal{A}_\tau |\mathcal{A}_e|}{\mathcal{A}}
    \end{matrix}\right]\left[ \begin{matrix} \nu_{e}\cr \nu_\mu\cr \nu_\tau \end{matrix}\right] =U\cdot \left[ \begin{matrix} \nu_{e}\cr \nu_\mu\cr \nu_\tau \end{matrix}\right],
\end{align}
where $\mathcal{A}=\sqrt{|\mathcal{A}_\tau|^2+|\mathcal{A}_\mu|^2}.$
Similarly, we can define the basis $(\overline{\nu}_{\rm far},\overline{\nu}_\perp, \overline{\nu}_T)$ replacing $U \to \overline{U}$ in which $\mathcal{A}_\alpha \to \overline{\mathcal{A}}_\alpha$.
In the new basis, we can write the couplings as 
\begin{align*}
    f^{Vq}_{\alpha \beta} \to \tilde{f}^{Vq}_{\alpha \beta}=(U\cdot f^{Vq}\cdot U^\dagger)_{\alpha \beta}=f^{Vq}_{\alpha \beta} \quad {\rm and} \quad f^{Aq}_{\alpha \beta} \to \tilde{f}^{Aq}_{\alpha \beta}=(U\cdot f^{Aq}\cdot U^\dagger)_{\alpha \beta}\ne f^{Aq}_{\alpha \beta}.
\end{align*}
Notice  that since $f^{Vq}\propto \delta_{\alpha \beta}$, $\tilde{f}^{Vq}=f^{Vq}.$ For antineutrinos, we should replace $U \to \overline{U}.$
A DUNE-like detector does not detect final neutrino in the neutrino nucleus scattering so the number of DIS NC events at the far detector are given by
\begin{align*}
    (\sigma_{n/p})_{\nu_{\rm far}} = \sum_{\alpha \in \{{\rm far},\perp, T\}} \sigma_{n/p}(\nu_{\rm far}+N \to \nu_\alpha +X),
\end{align*}
in the neutrino mode and 
\begin{align*}
    (\sigma_{n/p})_{\overline{\nu}_{\rm far}} =\sum_{\alpha \in \{{\rm far},\perp, T\}} \sigma_{n/p}(\overline{\nu}_{\rm far}+N \to \overline{\nu}_\alpha +X),
\end{align*}
in the antineutrino mode. These cross sections can be computed using Eq.~(\ref{eq:sigtot}), replacing $f^{Aq}\to \tilde{f}^{Aq}$. Remember that $\tilde{f}^{Aq}$ for neutrinos and antineutrinos are different.

\subsection{Resonance scattering}
\label{sec:res}
Below an energy of $\sim$3 GeV, the resonance scattering can be dominant. In such interactions, a nucleon inside the target is knocked off and scattered up to a resonant baryon ($\Delta$ or $N^*$) which then decays back to a pion and proton and neutron: $\nu+N\to \nu +(\Delta \ {\rm or} \ N^*)\to \nu +\pi +N$. Neutral pion will decay to a photon pair and start an electromagnetic shower. As discussed in \cite{Tingey:2022evd} invoking neural networks, it is in principle possible to differentiate resonance NC scattering from NC DIS. The axial vector coupling can only contribute to the resonances with odd parity.
The amplitudes of the relevant processes can be written as 
\begin{align}
    &\frac{\mathcal{M}(\nu_\alpha +p \to \nu_\alpha +{\rm Res})}{\mathcal{M}^{\rm SM}(\nu_\alpha +p \to \nu_\alpha +{\rm Res})}=\frac{1/2 +2 \epsilon^{Au}_{\alpha \alpha}+\epsilon^{Ad}_{\alpha \alpha}}{1/2},\notag \\
    &\left.\frac{\mathcal{M}(\nu_\alpha +p \to \nu_\beta +{\rm Res})}{\mathcal{M}^{\rm SM}(\nu_\alpha +p \to \nu_\alpha +{\rm Res})}\right|_{\alpha \ne \beta}= \frac{2 \epsilon^{Au}_{\alpha \beta}+\epsilon^{Ad}_{\alpha \beta}}{1/2},\notag \\
    &\frac{\mathcal{M}(\nu_\alpha +n \to \nu_\alpha +{\rm Res})}{\mathcal{M}^{\rm SM}(\nu_\alpha +n \to \nu_\alpha +{\rm Res})}= \frac{-1/2 + \epsilon^{Au}_{\alpha \alpha}+2\epsilon^{Ad}_{\alpha \alpha}}{-1/2}, \notag \\
    &\left.\frac{\mathcal{M}(\nu_\alpha +n \to \nu_\beta +{\rm Res})}{\mathcal{M}^{\rm SM}(\nu_\alpha +n \to \nu_\alpha +{\rm Res})}\right|_{\alpha \ne \beta}= \frac{ \epsilon^{Au}_{\alpha \beta}+2\epsilon^{Ad}_{\alpha \beta}}{-1/2},
\end{align}
where Res denotes $\Delta$ or $N^*$ resonant particle.
In this paper, we only focus on NC DIS and postpone discussing resonance events to a future work.

\section{NC NSI events at near and far detectors}
\label{sec:NCNSI}
The number of NC DIS at the Near Detector (ND) and at the Far Detector (FD) for the neutrino and antineutrino modes can be written as
\begin{align}
    \mathcal{N}^{\rm ND}_{{\nu}} &= \int \phi^{\rm ND}_{{\nu}} (E) \left[(\sigma_n)_{\nu_{\mu}} N_n^{\rm ND} + (\sigma_p)_{\nu_{\mu}} N_p^{\rm ND}\right] dE, \notag \\
    \mathcal{N}^{\rm ND}_{\overline{\nu}} &= \int \phi^{\rm ND}_{\overline{\nu}} (E) \left[(\sigma_n)_{\overline{\nu}_{\mu}} N_n^{\rm ND} + (\sigma_p)_{\overline{\nu}_{\mu}} N_p^{\rm ND}\right] dE, \notag \\
    \mathcal{N}^{\rm FD}_{\nu} &= \int \phi^{FD}_{\nu}(E) \left[(\sigma_n)_{\nu_{\rm far}} N_n^{\rm FD} + (\sigma_p)_{\nu_{\rm far}} N_p^{\rm FD}\right] dE, \notag \\
    \mathcal{N}^{\rm FD}_{\overline{\nu}} &= \int \phi^{\rm FD}_{\overline{\nu}}(E) \left[(\sigma_n)_{\overline{\nu}_{\rm far}} N_n^{\rm FD} + (\sigma_p)_{\overline{\nu}_{\rm far}} N_p^{\rm FD}\right] dE,
    \label{eq:number-of-events}
\end{align}
where $\phi^{\rm FD/ND}_{\nu/\overline{\nu}}$ are the time-integrated fluxes of neutrinos or antineutrinos at ND or FD in the absence of oscillation. The un-oscillated fluxes are given in Ref. \cite{noauthor_dune_nodate} per GeV/${\rm cm}^2/$POT per year. We have used the predicted $\phi^{\rm ND}_{\nu/\overline{\nu}}$ for the on-axis setup and have taken POT per year$ =1.1\times 10^{21}$ \cite{noauthor_dune_nodate} and 6.5 years for running in each neutrino and antineutrino mode as suggested by TDR \cite{dunecollaboration2021experiment}.
We have examined the results for both the so-called ``CP-optimized" and ``$\tau$-optimized" flux modes. 

${N}_{n/p}^{\rm ND/FD}$ are the numbers of protons and neutrons at the ND or at FD which are proportional to their fiducial masses $M_{\rm fid}^{\rm ND}=67.2\ {\rm ton}$ and $M_{\rm fid}^{\rm FD} = 40\ {\rm kton}$ \cite{DUNE:2020ypp,dunecollaboration2021experiment}.
We have considered the LAr detector at the near site. Including the GAr detector and SAND would yield slightly more statistics and therefore better results. In this sense, our results will be a bit conservative.
Considering that the detectors are made of Argon with 18 protons and 22 neutrons, we can write
\begin{align}
    N_p^{\rm ND/FD} = \frac{18}{40} \frac{M_{\rm fid}^{\rm ND/FD}}{M_p} \quad {\rm and} \quad N_n^{\rm ND/FD} = \frac{22}{40} \frac{M_{\rm fid}^{\rm ND/FD}}{M_p}.
\end{align}
Notice that the tiny difference between neutron and proton masses as well as the binding energy of Argon are neglected.
As we discussed in sect.~\ref{sec:DIS}, the cross sections $(\sigma_N)_{\nu_{\rm far}}$ and $(\sigma_N)_{\overline{\nu}_{\rm far}}$ depend on the oscillation amplitudes $\mathcal{A}_\alpha$ or $\overline{\mathcal{A}}_\alpha$.
Importing the details of the DUNE-like experiment represented in the ancillary files of Ref. \cite{dunecollaboration2021experiment} into the GLoBES 3.2.18 software \cite{Huber:2004ka,Huber:2007ji}, we have obtained the oscillation amplitudes. We have taken a constant matter density of $2.848\ {\rm g/cm^3}$ along the baseline \cite{dunecollaboration2021experiment,Roe:2017zdw}. For the neutrino mixing parameters, we invoke the values presented in NuFIT 5.2 (NO, w/o SK-atm) \cite{Esteban_2020} as shown in Tab. \ref{tab:mixing-parameters}. 
\begin{table}
    \caption{Three flavor neutrino oscillation parameters for Normal mass Ordering (NO) taken from NuFIT 5.2 global fit \cite{Esteban_2020}. The parameters have been obtained without inclusion of data on atmospheric neutrinos provided by the Super-Kamiokande collaboration (SK-atm).\label{tab:mixing-parameters}}
    \begin{ruledtabular}
        \begin{tabular}{c c c c c c}
            $\theta_{12}/\degree$ & $\theta_{23}/\degree$ &  $\theta_{13}/\degree$ & $\delta/\degree$ & $\Delta m_{12}^2/10^{-5} {\rm eV^2}$ & $\Delta m_{31}^2/10^{-3} {\rm eV^2}$ \\
            \colrule
            $33.41$ & $49.1$ & $8.54$ & $197$ & $7.41$ & $2.511$ \\
        \end{tabular}
    \end{ruledtabular}
\end{table}

Let us define
\begin{align}
    \mathcal{N}^{\rm ND}&\equiv \mathcal{N}^{\rm ND}_\nu + \mathcal{N}^{\rm ND}_{\overline{\nu}} \quad {\rm and} \quad \Delta \mathcal{N}^{\rm ND}\equiv\mathcal{N}^{\rm ND}_\nu - \mathcal{N}^{\rm ND}_{\overline{\nu}},\\
    \mathcal{N}^{\rm FD}&\equiv\mathcal{N}^{\rm FD}_\nu + \mathcal{N}^{\rm FD}_{\overline{\nu}} \quad  {\rm and} \quad \Delta \mathcal{N}^{\rm FD}\equiv\mathcal{N}^{\rm FD}_\nu - \mathcal{N}^{\rm FD}_{\overline{\nu}}.
\end{align}
Notice that these quantities depend on $\epsilon^{Af}$ through the dependence of the cross sections. At $\epsilon^{Af}=0$, these quantities approach the SM values:
\begin{align}
    \mathcal{N}^{\rm ND, FD}_{\rm SM}\equiv \mathcal{N}^{\rm ND,FD}(\epsilon^{Aq}_{\alpha \beta}=0) \quad {\rm and} \quad  \Delta\mathcal{N}^{\rm ND, FD}_{\rm SM}\equiv \Delta \mathcal{N}^{\rm ND,FD}(\epsilon^{Aq}_{\alpha \beta}=0).
\end{align}

\begin{figure}
    \centering
    \includegraphics[height=0.82\textheight]{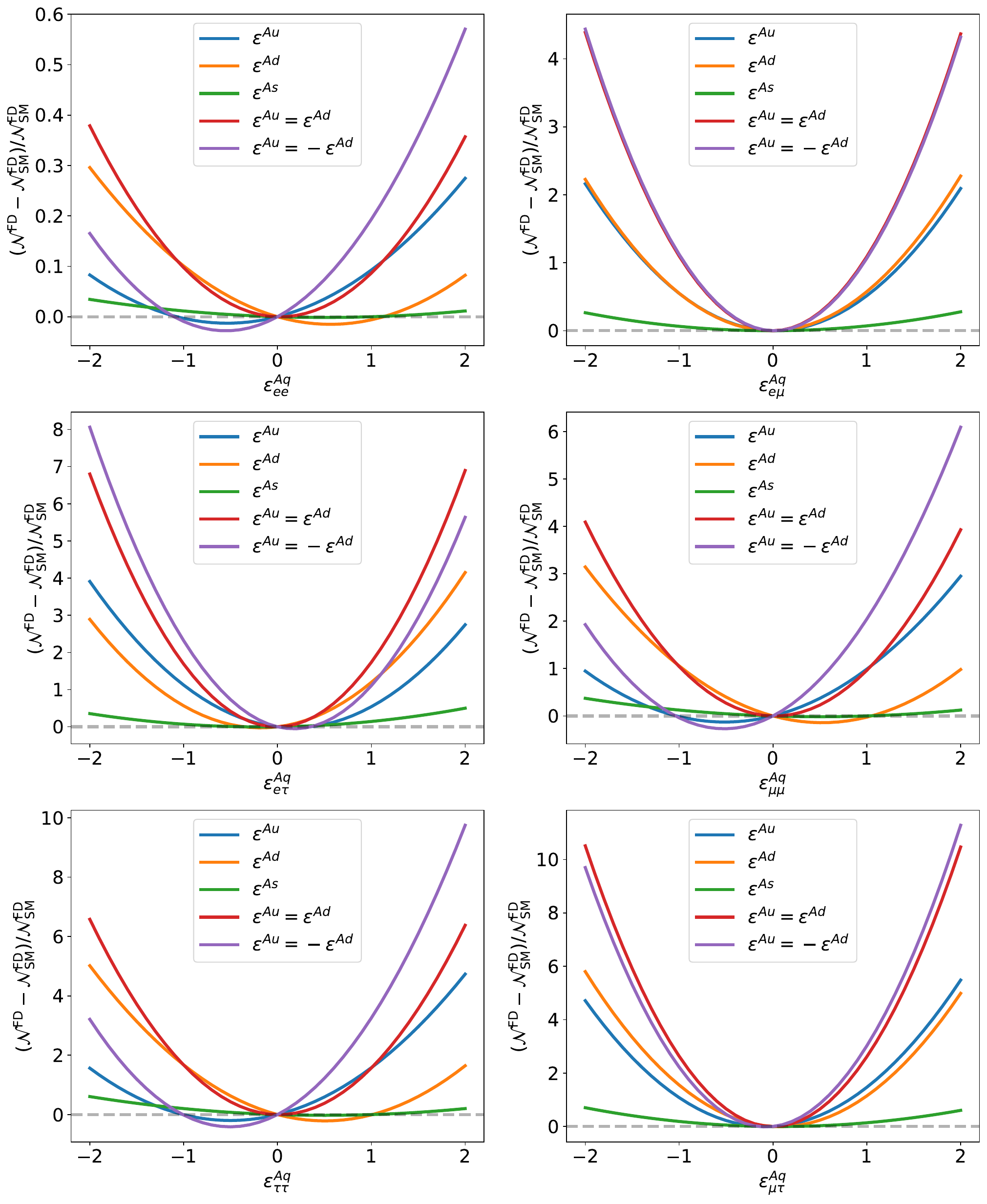}
    \caption{Deviation of the total number of NC DIS neutrino plus antineutrino events from the SM prediction at the far detector versus the NSI parameters. We have assumed equal data taking time with $1.1\times 10^{21}$ POT per year in each of neutrino and antineutrino modes and have taken the CP-optimized un-oscillated spectra as given in \cite{noauthor_dune_nodate}. For each curve, the $\epsilon^{Aq}$ elements that are not indicated in the legend are set to zero. All the elements are taken to be real.}
    \label{fig:NFD}
\end{figure}
\begin{figure}
    \centering
    \includegraphics[height=0.82\textheight]{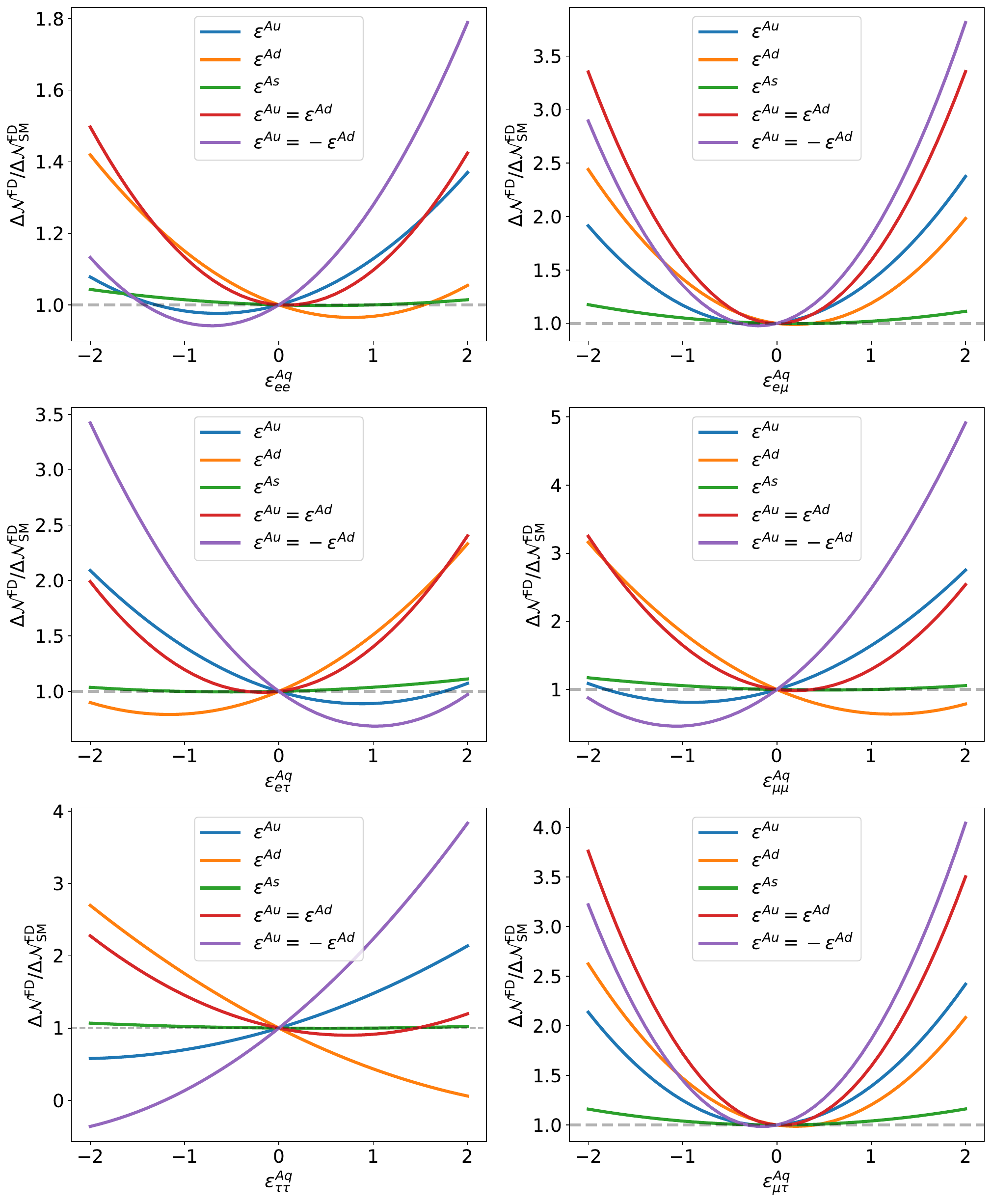}
    \caption{Ratio of the difference of the number of NC DIS events in the neutrino and antineutrino modes in the presence of NSI at the far detector to the SM prediction for the same difference versus the NSI parameters.
    We have assumed equal data taking time with $1.1\times 10^{21}$ POT per year in each of neutrino and antineutrino modes and have taken the CP-optimized un-oscillated spectra as given in \cite{noauthor_dune_nodate}. For each curve, the $\epsilon^{Aq}$ elements that are not indicated in the legend are set to zero. All the elements are taken to be real.}
    \label{fig:DeltaNFD}
\end{figure}
\begin{figure}
    \centering
    \includegraphics[height=0.82\textheight]{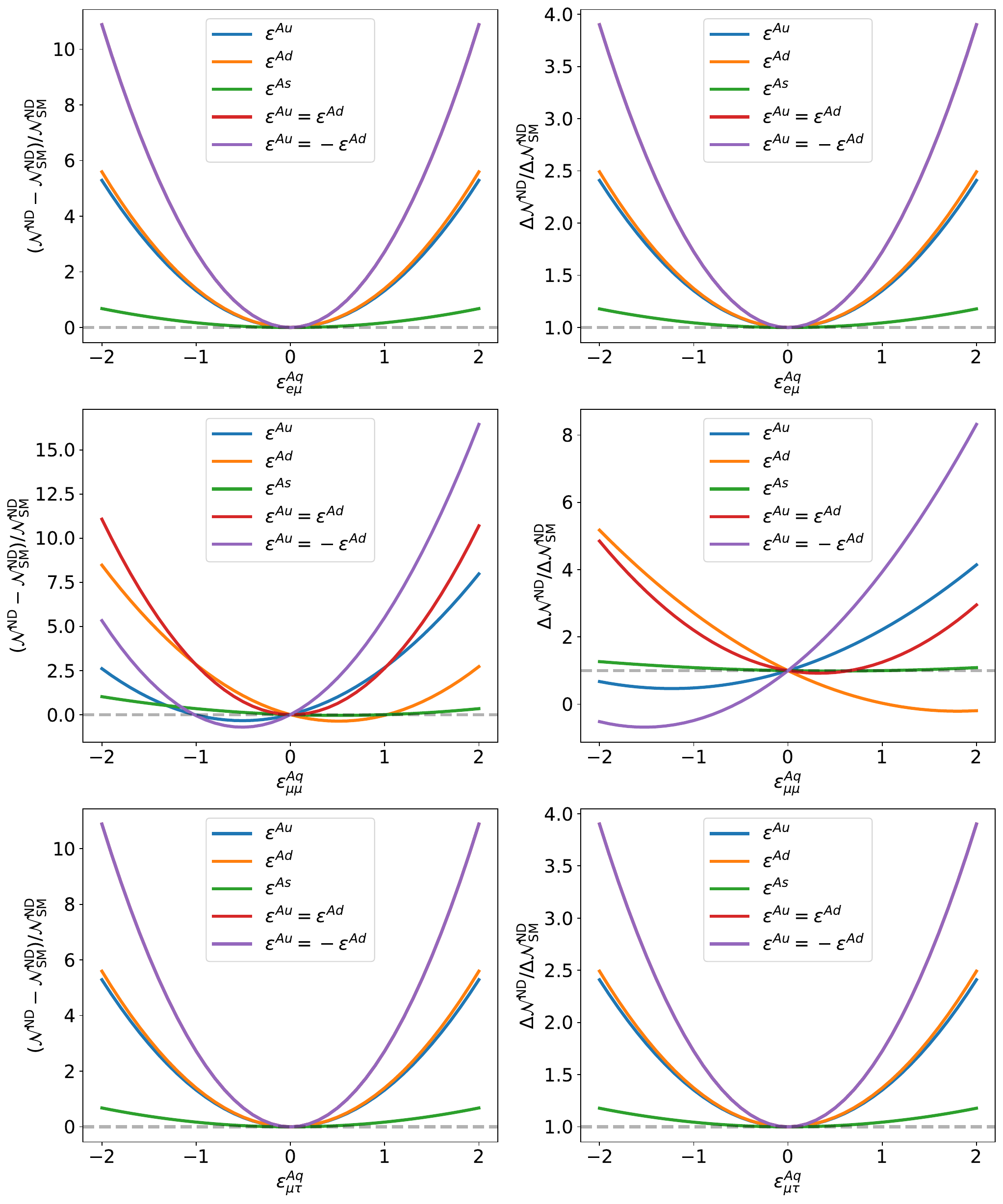}
    \caption{Left column: The same as Fig. \ref{fig:NFD} for near detector; Right column: The same as Fig. \ref{fig:DeltaNFD} for near detector. Since $\epsilon_{ee}^{Aq}$, $\epsilon_{e\tau}^{Aq}$ and $\epsilon_{\tau \tau}^{Aq}$ do not affect the events at the near detector, we do not include panels for them in this figure. All the elements are taken to be real.}
    \label{fig:NDtogether}
\end{figure}

Fig. \ref{fig:NFD} shows relative deviation of $N^{\rm FD}$ from $N^{\rm FD}_{\rm SM}$ as functions of various $\epsilon_{\alpha \beta}^{Aq}$. The legends in the panels show which NSI parameters are set nonzero for each curve. All curves reach zero at $\epsilon^{Aq}_{\alpha \beta}$ where SM is preserved. 
As the green curve in each panel shows, the sensitivity to $\epsilon^{As}$ is of order of one tenth of the sensitivity to $\epsilon^{Au}$. This is understandable because the contribution from the $s$-quark to the proton is about one tenth of that from the $u$-quark. That is $\int_0^1 dx x s(x)/\int_0^1 dx x u(x)\sim 0.1$ as shown in Table \ref{tab:moments}.
While sensitivities to the $e\mu$, $\mu\mu$, $\tau\tau$, $\tau \mu$ and $e\tau$ entries are comparable, the sensitivity to the $ee$ entry is smaller by a factor of order of 10.
This is because the $\nu_e$ component of $\nu_{\rm far}$ is smaller than the $\nu_\tau$ and $\nu_\mu$ components and suppressed by $\sin \theta_{13}$. Generally for the $e\mu$, $\mu\mu$, $\tau\tau$, $\tau \mu$ and $e\tau$ entries, the variation is of order of $\epsilon^{Aq}$, itself.

Even with equal POT per year at the neutrino and antineutrino modes,  $\mathcal{N}^{\rm FD,ND}_\nu$ will be different from $\mathcal{N}^{\rm FD,ND}_{\overline{\nu}}$; {\it i.e.,} $\Delta \mathcal{N}^{\rm FD,ND}_{\rm SM}\ne 0$. This is because the energy spectra of neutrinos and antineutrinos as well as their cross sections are different. The dependence of $\mathcal{N}^{\rm FD,ND}_\nu$ on NSI differs from that of $\mathcal{N}^{\rm FD,ND}_{\overline{\nu}}$. As a result, for $\epsilon^{Aq}\ne 0$, $\Delta \mathcal{N}^{\rm FD}/\Delta \mathcal{N}^{\rm FD}_{\rm SM} \ne 1$. Fig. \ref{fig:DeltaNFD} shows this ratio versus various $\epsilon_{\alpha \beta}^{Aq}$ elements. In the presence of NSI, this ratio may depend on the mixing parameters and in particular on $\delta$. We have examined this dependence and have found that the change with varying $\delta$ is negligible for $\epsilon_{\alpha \beta}^{Aq}<1.$

At the near detector, the neutrino beam is mainly composed of $\nu_\mu$ or $\overline{\nu}_\mu$ so there is no sensitivity to $\epsilon_{ee}^{Aq}$, $\epsilon_{e\tau}^{Aq}$ and $\epsilon_{\tau\tau}^{Aq}$. In Fig. \ref{fig:NDtogether}, we have therefore shown only the $\epsilon_{\mu\mu}^{Aq}$, $\epsilon_{\mu\tau}^{Aq}$ and $\epsilon_{\mu e}^{Aq}$ plots. The left panels show the relative deviation from the SM prediction and the right panels show $\Delta \mathcal{N}^{\rm ND}/\Delta \mathcal{N}^{\rm ND}_{\rm SM}$. The overall behavior is similar to what we have seen for the far detector. However, since at the ND, the neutrino flux and therefore the number of NC DIS events are much higher, the ND data will yield stronger bounds on $\epsilon_{\mu\mu}^{Aq}$, $\epsilon_{\mu\tau}^{Aq}$ and $\epsilon_{\mu e}^{Aq}$.
\section{Forecasting the bounds on axial NSI}
\label{sec:forecast}
In the previous section, we studied the dependence of the deviation of the number of events on NSI for different combinations of $\epsilon^{Au}$, $\epsilon^{Ad}$ and $\epsilon^{As}$. In this section, to compare our forecast  with the existing constraints discussed in sect.~\ref{CHARM-NuTeV}, we study the cases that only $\epsilon^{Au}$ or $\epsilon^{Ad}$ is non-zero.
We also focus on the benchmark point $\epsilon^{Au}/\epsilon^{Ad}=1$ for two reasons: (1) This benchmark is unconstrained by SNO; (2) For $\epsilon^{Au}=\epsilon^{Ad}$, the isospin symmetry is preserved which from the model building point of view is favored. We will also consider the possibility of testing the non-trivial SNO solution  ($\epsilon_{\tau \tau}^{Au}- \epsilon_{\tau \tau}^{Ad}\simeq-1.5$)  which is still allowed for an arbitrary ratio of $\epsilon^{Au}/\epsilon^{Ad}$. Moreover, we shall forecast the bounds on $\epsilon^{As}$ from the ND and the FD of a DUNE-like experiment.

The CC events as well as the resonance neutrino interaction events may be misidentified as a signal for DIS NC interactions. The backgrounds to the neutrino quark NC DIS signal are mainly composed of misidentified neutrino quark total CC interactions plus misidentified  neutrino quark resonant NC interactions. Since the muons produced in the CC interactions of $\nu_\mu$ are quite distinctive, the probability of misidentifying them as NC event is expected to be low ({\it i.e.,} $O(10 \%)$). Due to oscillation, the neutrino flux  at the far detector will contain a $\nu_\tau$ component which, if energetic enough, can produce $\tau$ via CC interactions. In 65 \% of cases when $\tau$ decays hadronically, the probability of misidentifying it as a NC event is high. However, the total number of $\nu_\tau$ CC events will be less than 1 \%  of $\nu_\mu$ CC events \cite{MammenAbraham:2022xoc} so even if all of the events are misidentified, their impact in our results will be negligible. We can therefore write
\begin{align}
    \mathcal{B}^{\rm ND/FD}_{\nu/\overline{\nu}} = \epsilon_{\rm CC} (\mathcal{N}^{\rm ND/FD}_{\rm CC})_{\nu/\overline{\nu}} + \epsilon_{\rm Res} (\mathcal{N}_{\rm Res}^{\rm ND/FD})_{\nu/\overline{\nu}},
\end{align} 
where $(\mathcal{N}_{\rm CC}^{\rm ND/FD})_{\nu/\overline{\nu}}$ and $(\mathcal{N}_{\rm Res}^{\rm ND/FD})_{\nu/\overline{\nu}}$ are respectively the numbers of the CC and resonant interaction events at the near or  far detector during the neutrino or antineutrino running modes. The $\epsilon_{\rm CC}$ and $\epsilon_{\rm Res}$ coefficients are their respective efficiencies when cuts are applied to collect the NC DIS events.
Following 
\cite{Coloma:2017ptb}, we assume that the total CC background to NC DIS will be suppressed by $\epsilon_{\rm CC} \sim 10\%$.
To our best knowledge, there is still no dedicated study on distinguishing the NC DIS events from NC quasi-elastic and resonance events at a DUNE-like experiment.
However, for the CHIPS proposal with a simple water Cherenkov detector, it is shown that by invoking the neural network technique, the distinction between the two is possible \cite{Tingey:2022evd}. We assume that in the state-of-the-art LAr detector of DUNE-like experiments, the neural network technique will make it possible to suppress the background from NC quasi-elastic and resonance events to $\epsilon_{\rm Res} \sim 10\%$.
According to Fig. 4.25 of Ref. \cite{DUNE:2020ypp}, one can write $(\mathcal{N}_{\rm CC})^{\rm FD}_\nu = {71} \mathcal{N}^{\rm FD}_\nu/12$ and $(\mathcal{N}_{\rm Res})^{\rm FD}_\nu = 7 \mathcal{N}^{\rm FD}_\nu/12$ where $\mathcal{N}_\nu^{\rm FD}$ is the number of NC DIS neutrino events at the far detector. A similar relation holds for the antineutrino mode.
Since the technologies of FD and ND are similar, one can use these relations for ND, too. Table \ref{tab:numev} presents the Standard Model predictions for the number of neutral current events and their background over 6.5 years of data taking in neutrino and antineutrino modes at both ND and FD, considering both CP-optimized and $\tau$-optimized fluxes.

\begin{table}
    \caption{Standard Model prediction for the number of NC events and its background for 6.5 years of data taking in neutrino and antineutrino modes at ND and FD for both CP-optimized and $\tau$-optimized fluxes.
    \label{tab:numev}}
    \begin{ruledtabular}
        \begin{tabular}{c c c c c c}
            Mode & Flux & $\mathcal{N}^{\rm ND}$ & $\mathcal{B}^{\rm ND}$ & $\mathcal{N}^{\rm FD}$ & $\mathcal{B}^{\rm FD}$ \\
            \colrule
            \multirow{2}{-4em}{$\nu$}&CP     & 198334815 & 128917629 & 18835 & 12242 \\
            &$\tau$ & 528621779 & 343604156 & 46059 & 29938 \\
            \colrule
            \multirow{2}{-4em}{$\overline{\nu}$}&CP     & 88543625 & 57553356 & 8244 & 5358 \\
            &$\tau$ & 209747742 & 136336032 & 18145 & 11794 \\
        \end{tabular}
    \end{ruledtabular}
\end{table}

Let us first discuss the $\epsilon_{ee}^{Aq}$, $\epsilon_{e\tau }^{Aq}$ and $\epsilon_{\tau \tau}^{Aq}$ elements on which the present bounds are rather weak. Since $\epsilon^{Aq}$ is a Hermitian matrix, only its off-diagonal elements can be complex. We first study the case of real $\epsilon_{e \tau}^{Aq}$ but then forecast the bounds in the Arg($\epsilon^{Aq}_{e\tau}$) and $|\epsilon^{Aq}_{e\tau}|$ plane. As we saw in the previous section, if the $ \epsilon_{e\tau}^{Au/d}$ or $\epsilon_{\tau\tau}^{Au/d}$ elements vary in the allowed range, the deviation of number of events from the SM prediction in the far detector will be of the order of 1. Considering that we expect $\mathcal{O}(10^4)$ NC events at the FD of a DUNE-like experiment, statistics will allow determination of these couplings down to $\mathcal{O}(10^{-2})$. 
The important question is whether systematic errors can allow such precision. As seen from table~\ref{tab:moments}, the uncertainties induced from PDFs on the NC cross section calculation are below the percent level so these uncertainties will not be a serious limiting factor. A more serious source of uncertainty is the nuclear effects such as shadowing and anti-shadowing. The $\epsilon_{ee}^{Aq}$, $\epsilon_{e\tau}^{Aq}$ and $\epsilon_{\tau \tau}^{Aq}$ elements do not affect events at ND so in principle, ND with more than $\mathcal{O}(10^8)$ NC events and even larger sample of CC events can determine the flux of un-oscillated neutrinos times the SM cross sections with high precision. In the presence of NSI (but not within the flavor universal SM), the number of NC events at FD also depends on the mixing parameters which suffer from uncertainties. The uncertainty on $\Delta m_{31}^2$, $\sin^2 \theta_{23}$ and $\sin^2 \theta_{13}$ are already at the percent level. The dependence of the number of NC events at FD on the rest of the mixing parameters is sub-dominant so their uncertainty will not significantly affect  $\mathcal{N}^{\rm FD}$.
Anyway in the absence of NSI, both $\mathcal{N}^{\rm FD}_\nu$ and $\mathcal{N}^{\rm FD}_{\overline{\nu}}$ are independent of mixing parameters so a deviation from the SM prediction cannot be attributed to the uncertainties in the mixing parameters. That is the bound on $\epsilon^{Aq}$ will not suffer from the mixing parameter uncertainties.

For the purpose of forecasting the bounds on the NSI parameters, we define the statistics $\chi^2$ as follows

\begin{align}
		\chi^2 = \left[\sum_{Y = \nu, \overline{\nu}} \left( \frac{\left[\xi \mathcal{N}_Y^{\rm FD}(\epsilon_{\rm test}^{Aq})-\epsilon  \mathcal{N}_Y^{\rm FD}(\epsilon^{Aq}=0) +\omega_{Y}\mathcal{B}^{\rm FD}_{Y}\right]^2}{\epsilon \mathcal{N}_Y^{\rm FD}(\epsilon^{Aq}=0) + \mathcal{B}^{\rm FD}_Y} +\frac{ \omega_{Y}^2}{\sigma_{\omega}^2} \right) +\frac{(\xi -\epsilon)^2}{\sigma_\epsilon^2} \right]_{\rm min},\label{chiNOT}
\end{align}
where 
the minimization is made with respect to the pull parameters $\xi$, $\omega_{\nu}$, and $\omega_{\bar{\nu}}$.
Here, $\epsilon$ is the efficiency of detecting the signal which we take to be 90\% \cite{Coloma:2017ptb}.
We assume an uncertainty of $\sigma_\epsilon$ for $\epsilon$ and treat it with the pull parameter, $\xi$.   In fact, $\sigma_\epsilon$ takes care of the uncertainty in the prediction of the events that can come from the normalization of the un-oscillated neutrino flux, total cross section (mainly from shadowing, anti-shadowing and EMC effects) and the efficiency error estimation. The key point is that we take these uncertainties for the neutrino and antineutrino modes correlated and treat them with a single pull parameter $\xi$. As we shall discuss below, with separate pull parameters for the neutrino and antineutrino modes, some of the results would be dramatically different.
It is reasonable to assume that the uncertainty in the efficiency of the signal as well as the nuclear uncertainties of cross sections are the same for the neutrino and antineutrino modes so they can be treated by the same pull parameter $\xi$. The uncertainty in the normalization of the un-oscillated flux can be different for the neutrino and antineutrino modes but as discussed before, this uncertainty can be reduced by the measurement of the CC interaction at the ND of a DUNE-like experiment. 
The background from the CC events does not depend on the value of $\epsilon_{\alpha \beta}^{A f}$ so it does not appear in the numerator of the first term in Eq. (\ref{chiNOT}). However, as discussed in sect. \ref{sec:res}, the resonance DIS can also be affected by the value of $\epsilon_{\alpha \beta }^{Af}$. To be precise, we had to include $\epsilon^{\rm Res}_{\rm NC} \mathcal{N}^{\rm Res}_{\rm NC}(\epsilon_{\alpha \beta}^{A f})- \epsilon^{\rm Res}_{\rm NC} \mathcal{N}^{\rm Res}_{\rm NC}(\epsilon_{\alpha \beta}^{A f}=0)$.
Taking $\epsilon_{\rm NC}^{\rm Res}\sim 0.1$, we have neglected this contribution whose effect is expected to be less than 10 \%. 
However to account for the miscalculation of the backgrounds in the neutrino and antineutrino modes, we have included two separate pull  parameters $\omega_\nu$ and $\omega_{\bar{\nu}}$ with an uncertainty of $\sigma_\omega$.  Introducing two separate pull parameters for neutrino and antineutrino seems to be overly conservative because similarly to the case of efficiencies, the errors in the background estimation are expected to be correlated. We shall comment on the impact of correlated errors, too.
We take two sets of  nominal values for $\sigma_\epsilon$ and $\sigma_\omega$ with $\sigma_\epsilon=0,10\%$ and  $\sigma_\omega=0,2\%$ to study the dependence of the results on $\sigma_\epsilon$ and $\sigma_\omega$. For $\sigma_\epsilon=0$ (for $\sigma_\omega=0$), the minimum will lie at $\xi=\epsilon$  (at $\omega_\nu=\omega_{\bar{\nu}}=0$) so it is as if no pull parameter is introduced for the efficiency of the signal (for the background calculation).

In NC events, a sizeable fraction of the energy of the initial neutrinos are carried away by the final neutrinos so the deposited energy will be only a fraction of the initial energy. However, with the migration matrices, the deposited energy can be related to the initial neutrino energy \cite{Coloma:2017ptb,DeRomeri:2016qwo}.
With this method, the energy spectrum of NC DIS events can be studied. Such information can help to further suppress the background. Moreover, it can help to distinguish this model from other new physics, such as oscillation to sterile neutrinos \cite{Coloma:2017ptb} that can also affect the NC events but with a different energy dependence.
We do not however bin the spectrum and consider only the total number of NC DIS events to compute chi-square.

\begin{figure}
    \centering
    \includegraphics[height=0.82\textheight]{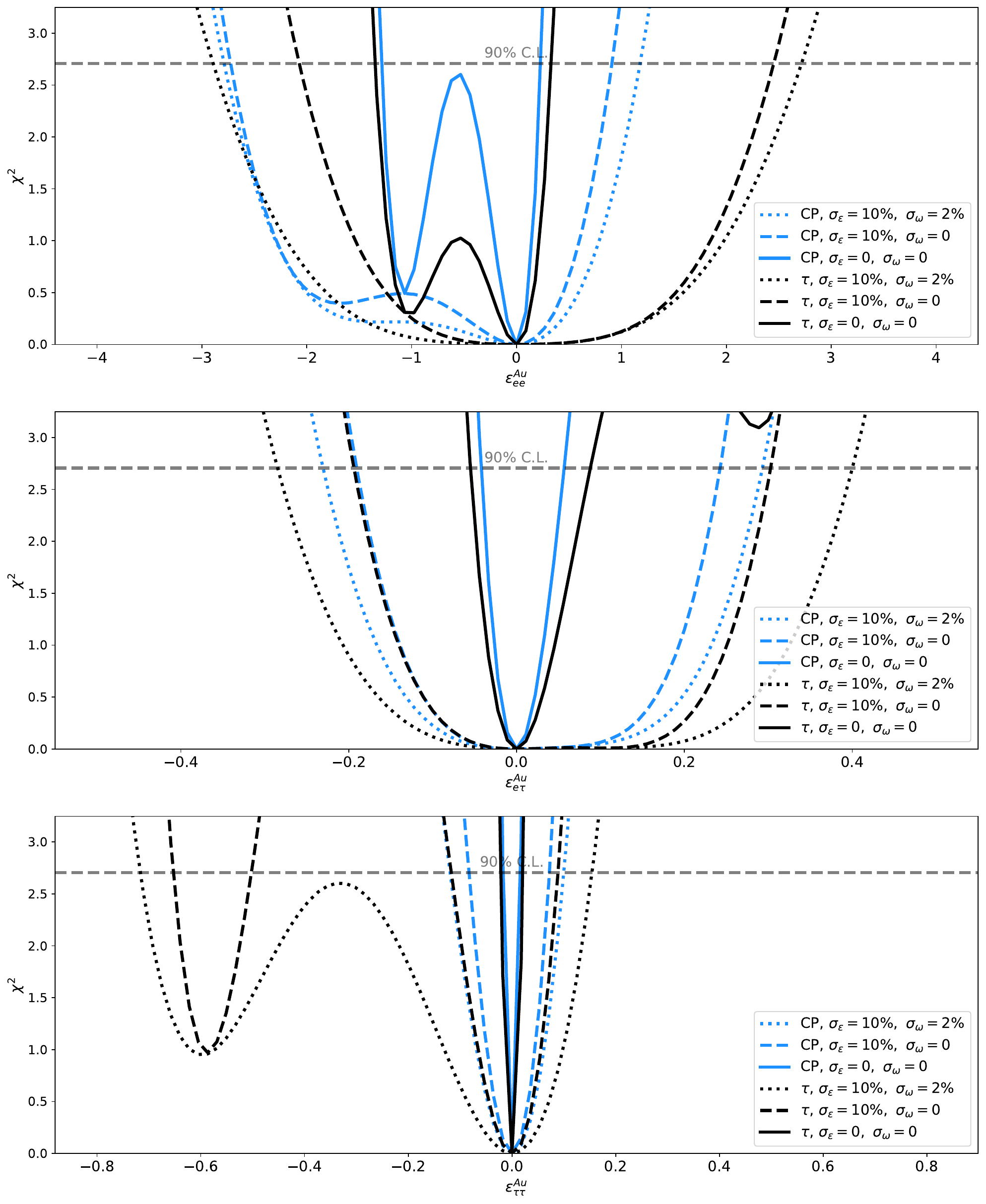}
    \caption{ 
    $\chi^2$ versus
    $\epsilon^{Au}$ for 6.5+6.5 years of data taking at FD. In each panel, only one flavor component is set nonzero.
    The intersections of the horizontal dashed line with the curves indicate the 90\% C.L. constraint. The blue (black) curves correspond to the CP-optimized ($\tau$-optimized) spectrum. For the solid curves, we have assumed zero uncertainty. For the dashed (dotted) curves, we have taken $\sigma_{\epsilon} =10\%$ and  $\sigma_{\omega} =0$ ($\sigma_{\epsilon} =10\%$ and $\sigma_{\omega} =2\%$).
    \label{fig:Chi2FD1u}}
\end{figure}

\begin{figure}
    \centering
\includegraphics[height=0.82\textheight]{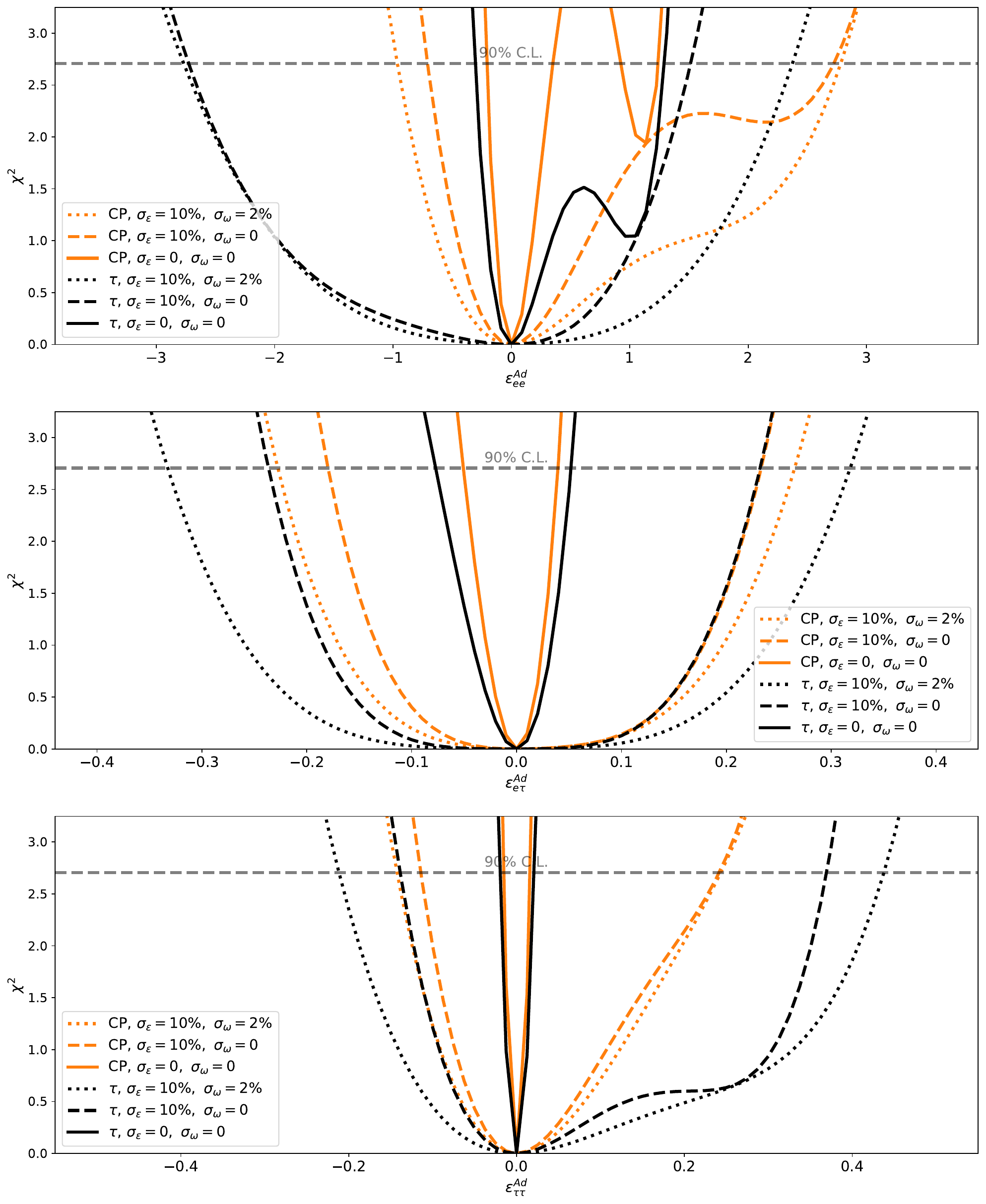}
    \caption{Similar to Fig.~\ref{fig:Chi2FD1u} for the axial NSI with the $d$ quark, $\epsilon^{Ad}$. The mustard (black) curves correspond to the CP-optimized ($\tau$-optimized) spectrum.\label{fig:Chi2FD1d}}
\end{figure}

\begin{figure}
    \centering
\includegraphics[height=0.82\textheight]{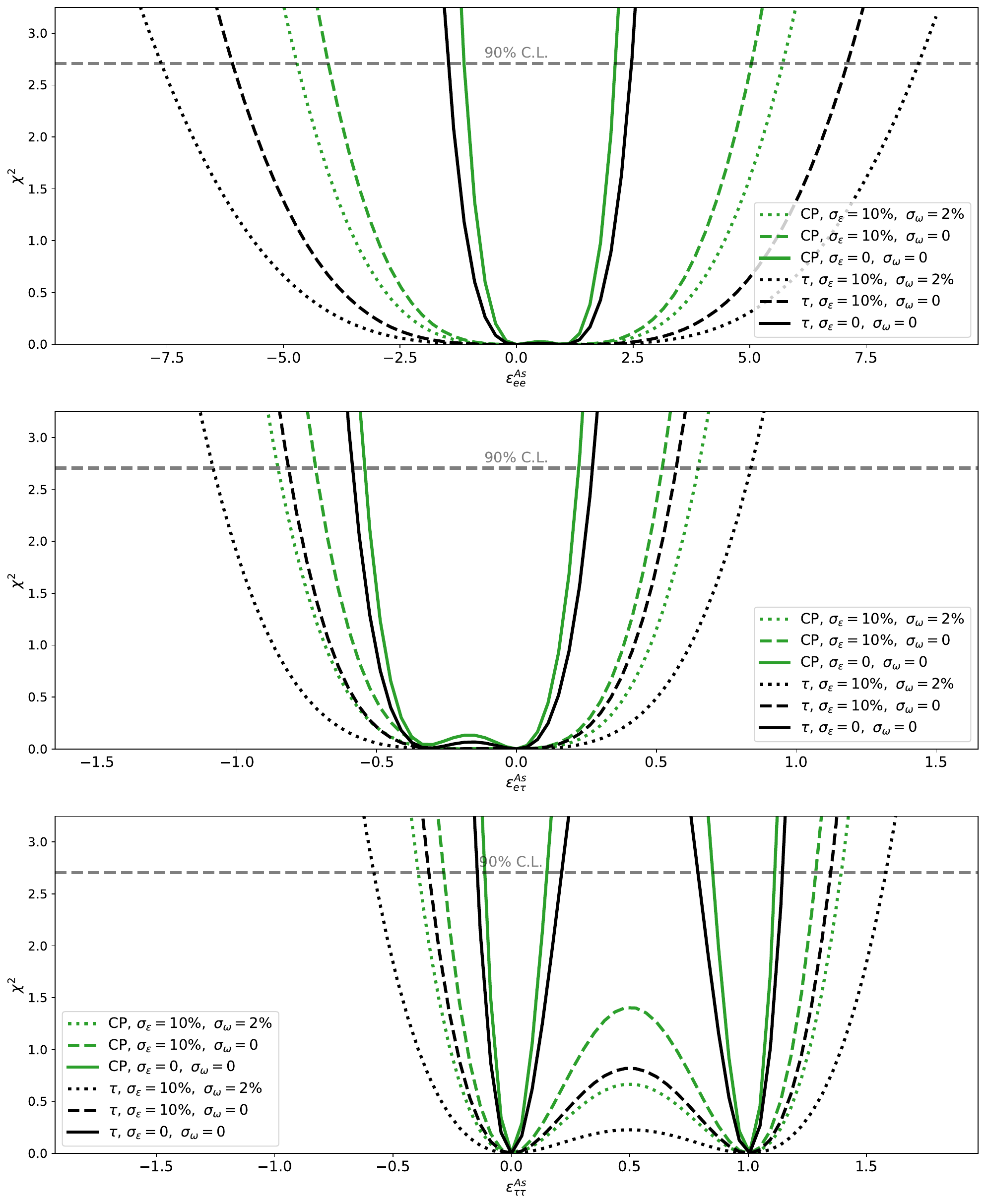}
    \caption{Similar to Fig.~\ref{fig:Chi2FD1u}  for the axial NSI with the $s$ quark, $\epsilon^{As}$. The green (black) curves correspond to the CP-optimized ($\tau$-optimized) spectrum. \label{fig:Chi2FD1s}}
\end{figure}

\begin{figure}
    \centering
    \includegraphics[height=0.82\textheight]{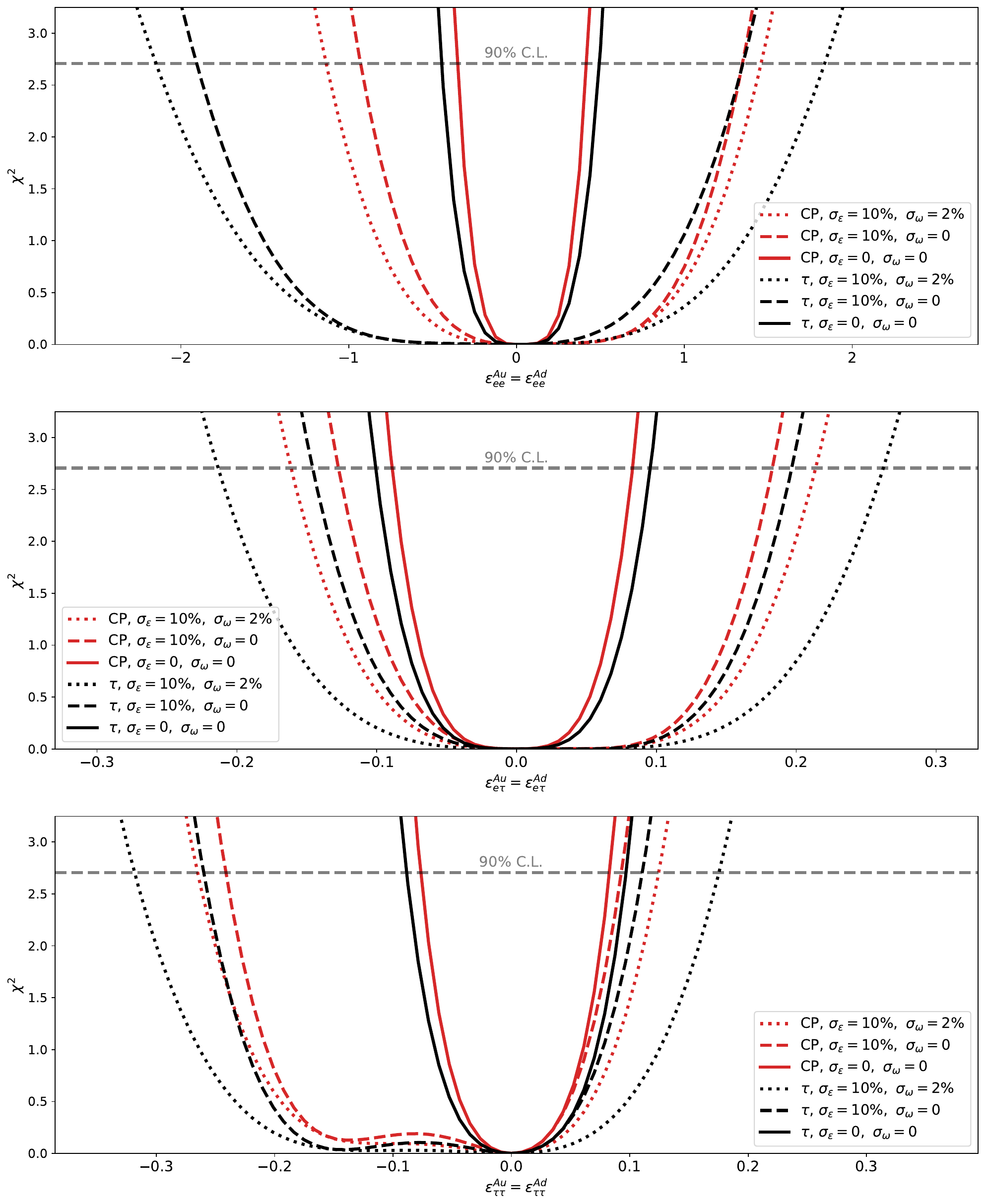}
    \caption{Similar to Fig.~\ref{fig:Chi2FD1u} except that the axial NSI with the $d$ and $u$ quarks are set to be equal, $\epsilon^{Ad}=\epsilon^{Au}$. The red (black) curves correspond to the CP-optimized ($\tau$-optimized) spectrum.\label{fig:Chi2FD1ueqd}}
\end{figure}

\begin{figure}
    \centering
    \includegraphics[height=0.27\textheight]{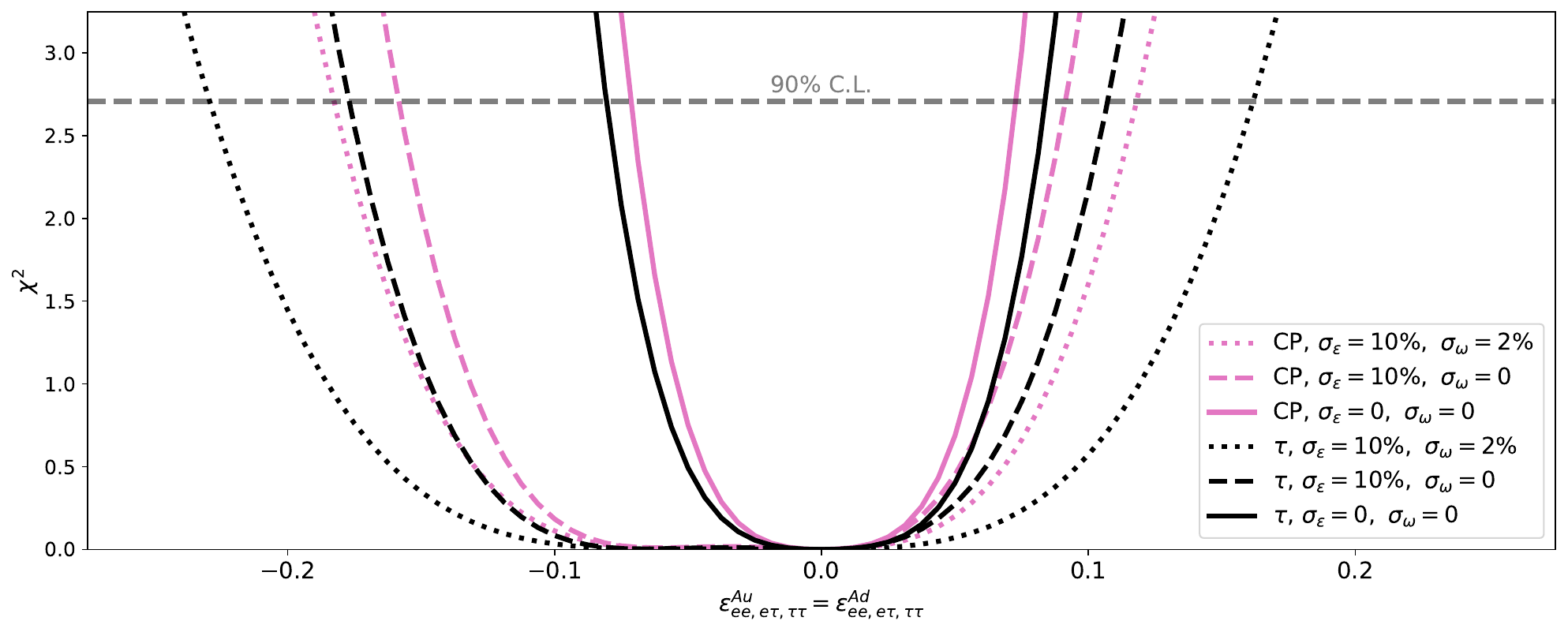}
    \caption{Similar to Fig.~\ref{fig:Chi2FD1ueqd} except that the $e\tau$ components of the axial NSI with the $d$ and $u$ quarks are set equal to the $ee$ and $\tau \tau$ components: $\epsilon^{Au}_{ee,e\tau,\tau\tau}=\epsilon^{Ad}_{ee,e\tau,\tau\tau}$. The pink (black) curves correspond to the CP-optimized ($\tau$-optimized) spectrum. \label{fig:Chi2FD3ueqd}}
\end{figure}

In Figs. \ref{fig:Chi2FD1u}-\ref{fig:Chi2FD3ueqd} we see $\chi^2$ for $\epsilon_{ee}^{Aq}$, $\epsilon_{\tau e}^{Aq}$ and $\epsilon_{\tau \tau}^{Aq}$.
The phase of the $e\tau$ element is set to zero.
The horizontal lines show $\chi^2$ corresponding to 90\% for each panel ($\chi^2=2.7$ for 1 d.o.f). 
In each panel of Figs. \ref{fig:Chi2FD1u}-\ref{fig:Chi2FD1s}, only one $\epsilon_{\alpha \beta}^{Aq}$ is taken to be nonzero but in Figs. \ref{fig:Chi2FD1ueqd} and \ref{fig:Chi2FD3ueqd}, $\epsilon^{Au}=\epsilon^{Ad}$. While in each panel of Figs. \ref{fig:Chi2FD1u}-\ref{fig:Chi2FD1ueqd} only one flavor entry of the $\epsilon_{\alpha \beta}^{Aq}$ matrix is nonzero, in Fig.~\ref{fig:Chi2FD3ueqd}, we have chosen a flavor structure that is favored from the model building point of view \cite{Farzan:2016wym}: $\epsilon^{Au/d}_{ee}=\epsilon^{Au/d}_{e\tau}=\epsilon^{Au/d}_{\tau\tau}$. 

As seen from the $ee$ panel of Fig.~\ref{fig:Chi2FD1u} along with a solution at $\epsilon_{ee}^{Au}=0$, there is another solution near $\epsilon_{ee}^{Au}=-1$. This is where $f^{Au}_{ee}\to - f^{Au}_{ee}$ at which the sum of cross sections of neutrinos and antineutrinos remains the same as when $\epsilon^{Aq}=0$ but the difference between neutrino and antineutrino cross sections partly solves the degeneracy (see the term in the fourth line of Eq.~(\ref{eq:sigtot})).
From Figs. \ref{fig:Chi2FD1d} and \ref{fig:Chi2FD1s}, we observe that such a degeneracy also exists for $\epsilon_{ee}^{Ad}\simeq 1$ and $\epsilon_{ee}^{As}\simeq 1$. Moreover, similar degeneracy exists at $\epsilon_{\tau \tau}^{As}=1$, at $\epsilon_{\tau \tau}^{Ad}=1$ and at $\epsilon_{\tau \tau}^{Au}=-1$. With the CP-optimized mode, the degeneracies at $\epsilon_{\tau \tau}^{Au}=-1$ and $\epsilon_{\tau \tau}^{Ad}=1$ are removed by neutrino antineutrino difference at high confidence level. (Notice that the fourth and fifth lines of Eq. (\ref{eq:sigtot}) are odd in $ f^{Aq}$ so the neutrino antineutrino difference can solve the degeneracy at $f^{Aq}\to -f^{Aq}$.) However, for $\epsilon_{\tau\tau}^{As}=1$ and for $\epsilon_{ee}^{As}=1$, the degeneracy is complete because $\int x^n (s-\overline{s}) dx=0$ and as a result the linear terms in $f^{As}$ which solve the degeneracy disappear.
Fig.~\ref{fig:Chi2FD1ueqd} demonstrates that there is no such degeneracy in the case $\epsilon_{ee}^{Au}=\epsilon_{ee}^{Ad}$. 

The solid curves in the figures show the results for zero uncertainty. The dashed curves correspond to  $\sigma_{\epsilon} =10\%$ and $\sigma_{\omega} =\omega_{\nu}=\omega_{\bar{\nu}}=0$. Finally, the dotted curves show the results with $\sigma_{\epsilon} =10\%$ and $\sigma_{\omega} =2\%$. As expected, the bounds become deteriorated when the uncertainties are taken to be nonzero.
Let us  focus on the impact of $\sigma_\epsilon$.
 For the elements that the sensitivity is high ({\it i.e.,} $\epsilon^{Au/d}_{e\tau}$ and $\epsilon^{Au/d}_{\tau\tau}$), the 90\% C.L. bound becomes of order of $\sigma_\epsilon$ itself but for those components that the sensitivity is low ({\it i.e.,} $\epsilon^{Au/d}_{ee}$ and $\epsilon^{As}$), the deterioration of the bound is more severe. In the figures, the black (colored) curve shows the results with $\tau$-optimized (CP-optimized) spectrum. As seen from the figures, the bound on $\epsilon_{ee}^{Aq}$ and $\epsilon_{\tau \tau}^{Aq}$ are expected to be stronger with a $\tau$-optimized spectrum. This is because at lower energies where the CP-optimized spectrum is peaked, the oscillation amplitudes $\mathcal{A}_e$ and $\mathcal{A}_\tau$ are larger than those at the higher energies of the $\tau$-optimized spectrum.

\begin{figure}
    \centering
    \includegraphics[height=0.29\textheight]{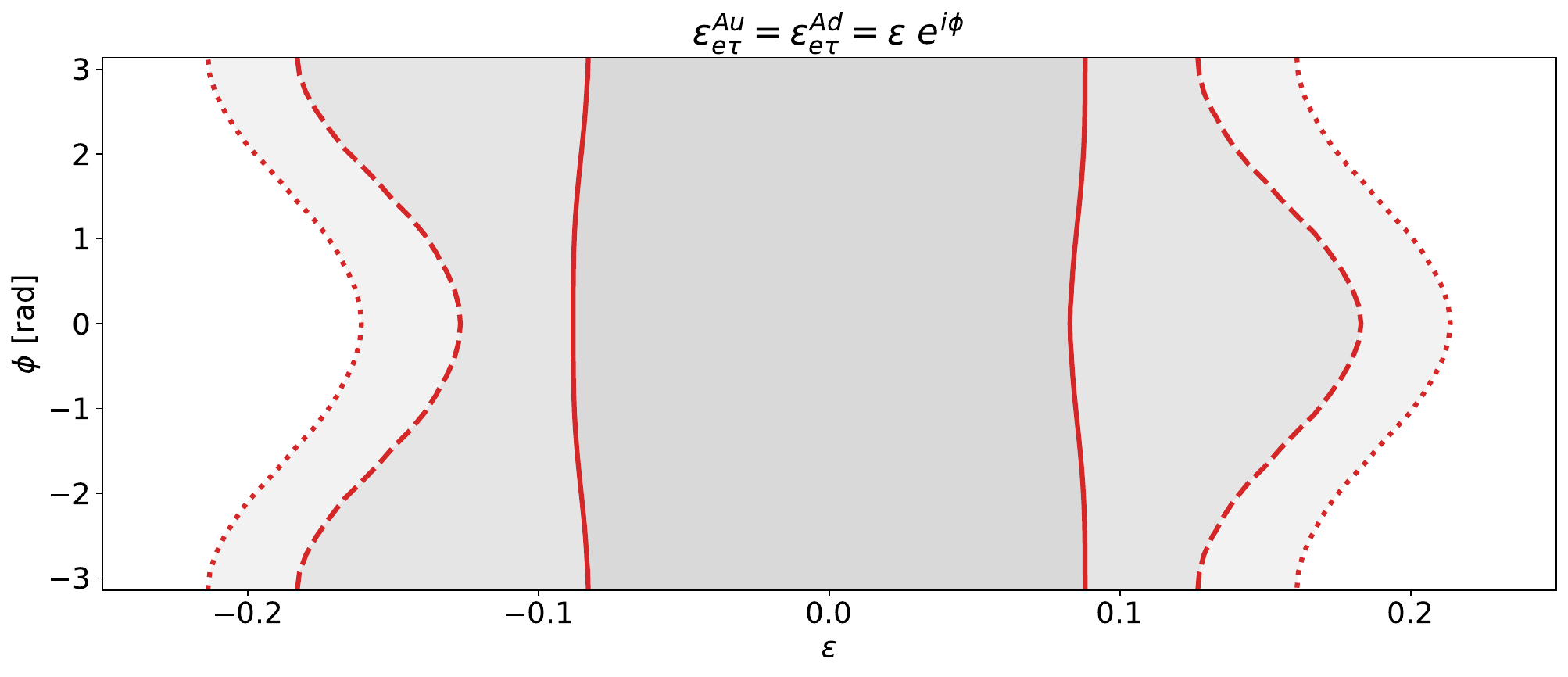}
    \caption{The 90\% C.L. bounds on the magnitude and argument of  $\epsilon_{e\tau}^{Au}=\epsilon_{e\tau}^{Ad}$ with the CP-optimized spectrum after 6.5+6.5 years of data taking. For the solid curves, we have assumed zero uncertainty. For the dashed (dotted) curves, we have taken $\sigma_{\epsilon} =10\%$ and  $\sigma_{\omega} =0$ ($\sigma_{\epsilon} =10\%$ and $\sigma_{\omega} =2\%$). \label{fig:Chi2FDContour1ueqd}}
\end{figure}
\begin{figure}
    \centering
    \includegraphics[height=0.27\textheight]{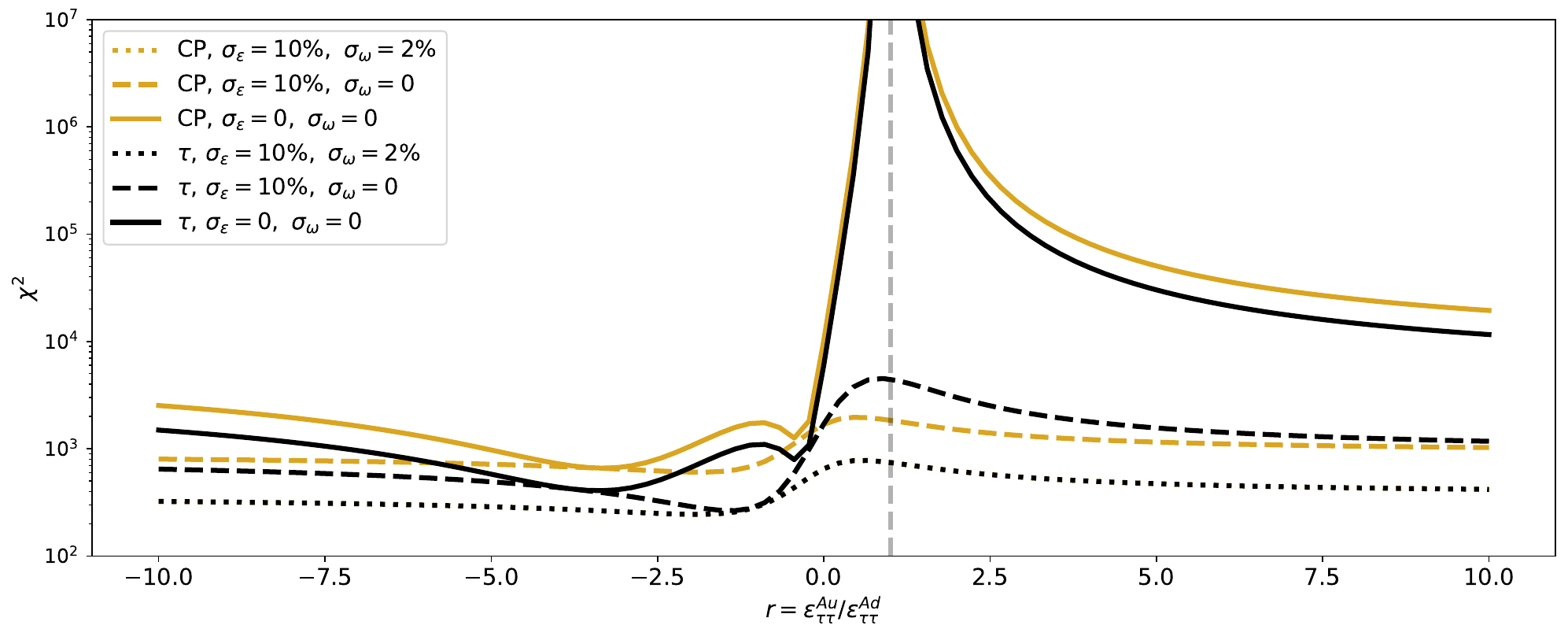} 
    \caption{$\chi^2$ after 6.5+6.5 years of data taking by FD of a DUNE-like experiment versus $r=\epsilon_{\tau \tau}^{Au}/\epsilon_{\tau \tau}^{Ad}$. The difference $\epsilon_{\tau \tau}^{Au}-\epsilon_{\tau \tau}^{Ad}$ is fixed to $ -1.5$ as indicated by the SNO solutions \cite{Coloma:2023ixt}.  The gold (black) curves correspond to the CP-optimized ($\tau$-optimized) spectrum. The golden dotted and the black dotted curves are not distinguishable. \label{fig:r}}
\end{figure}
\begin{figure}
    \centering
    \includegraphics[height=0.82\textheight]{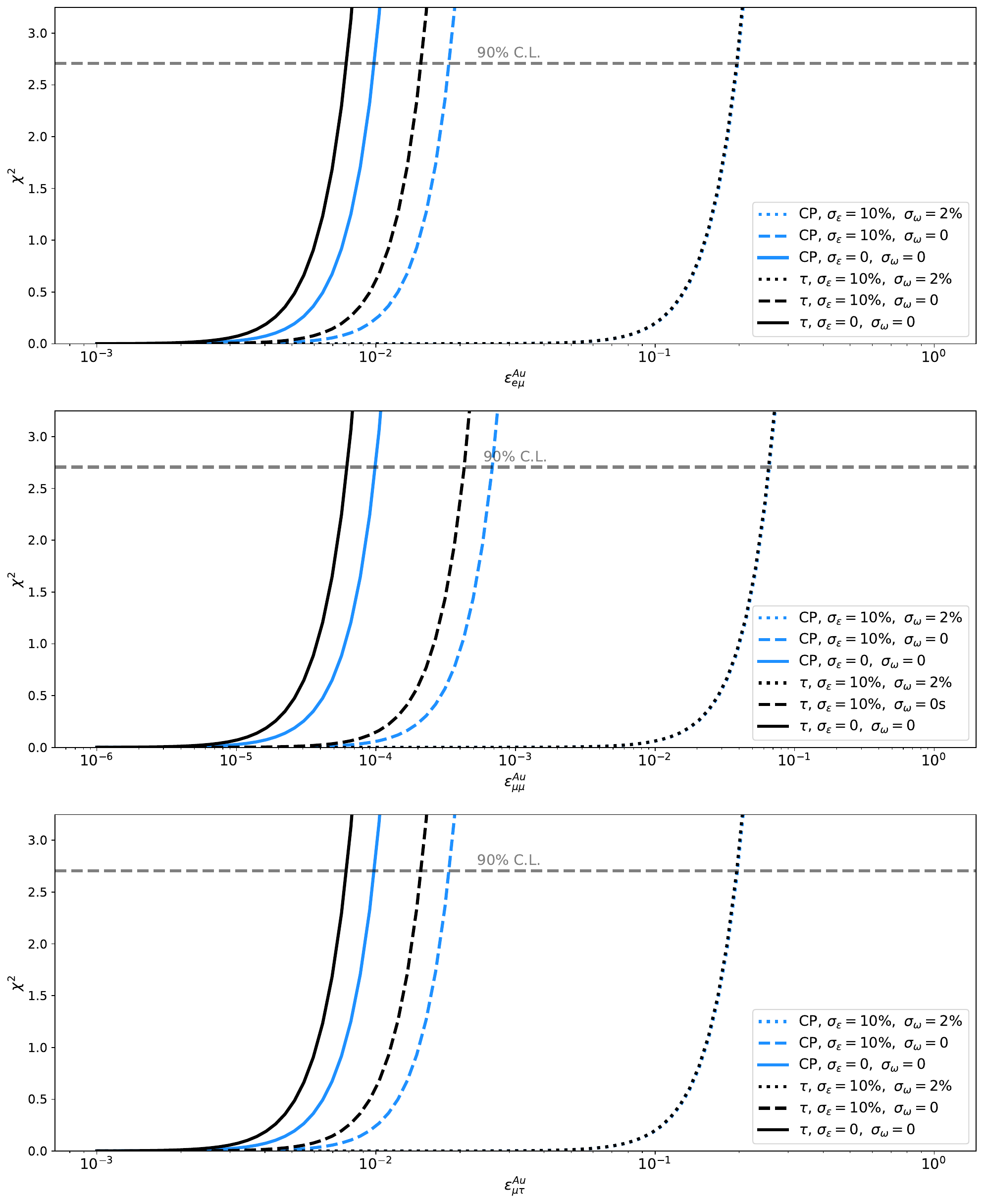}
    \caption{$\chi^2$ versus
    $\epsilon^{Au}$ for 6.5+6.5 years of data taking at ND. In each panel, only one flavor component is set nonzero.
    The intersections of the horizontal dashed line with the curves indicate the 90\% C.L. constraint. The blue (black) curves correspond to CP-optimized ($\tau$-optimized) spectrum. For the solid curves, we have assumed zero uncertainty. For the dashed (dotted) curves, we have taken $\sigma_{\epsilon} =10\%$ and  $\sigma_{\omega} =0$ ($\sigma_{\epsilon} =10\%$ and $\sigma_{\omega} =2\%$). \label{fig:Chi2ND1u}}
\end{figure}
\begin{figure}
    \centering
    \includegraphics[height=0.82\textheight]{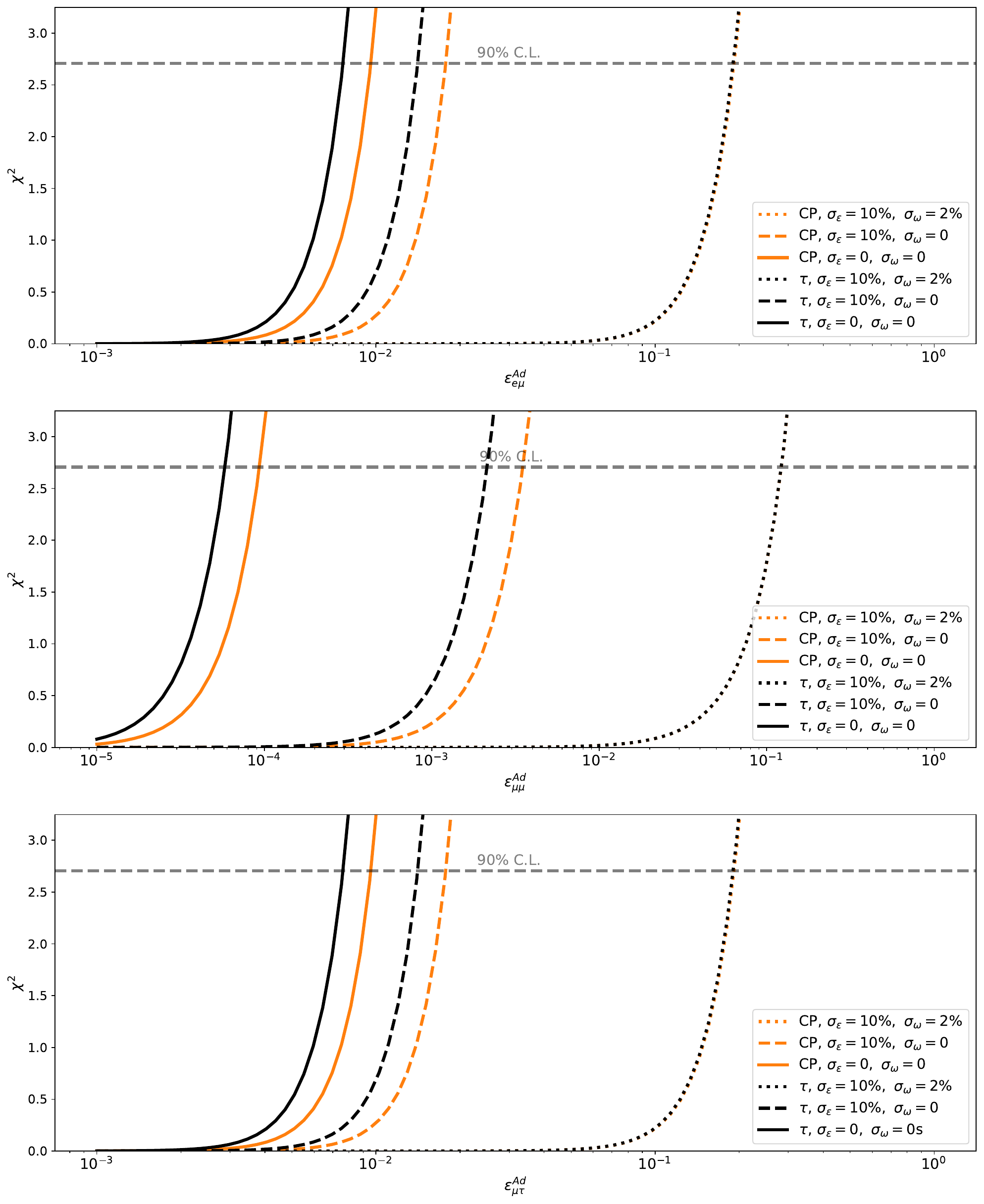}
    \caption{Similar to Fig.~\ref{fig:Chi2ND1u} for the axial NSI with the $d$ quark, $\epsilon^{Ad}$. The mustard (black) curves correspond to the CP-optimized ($\tau$-optimized) spectrum. The mustard dotted and the black dotted curves are similar. \label{fig:Chi2ND1d}}
\end{figure}
\begin{figure}
    \centering
    \includegraphics[height=0.82\textheight]{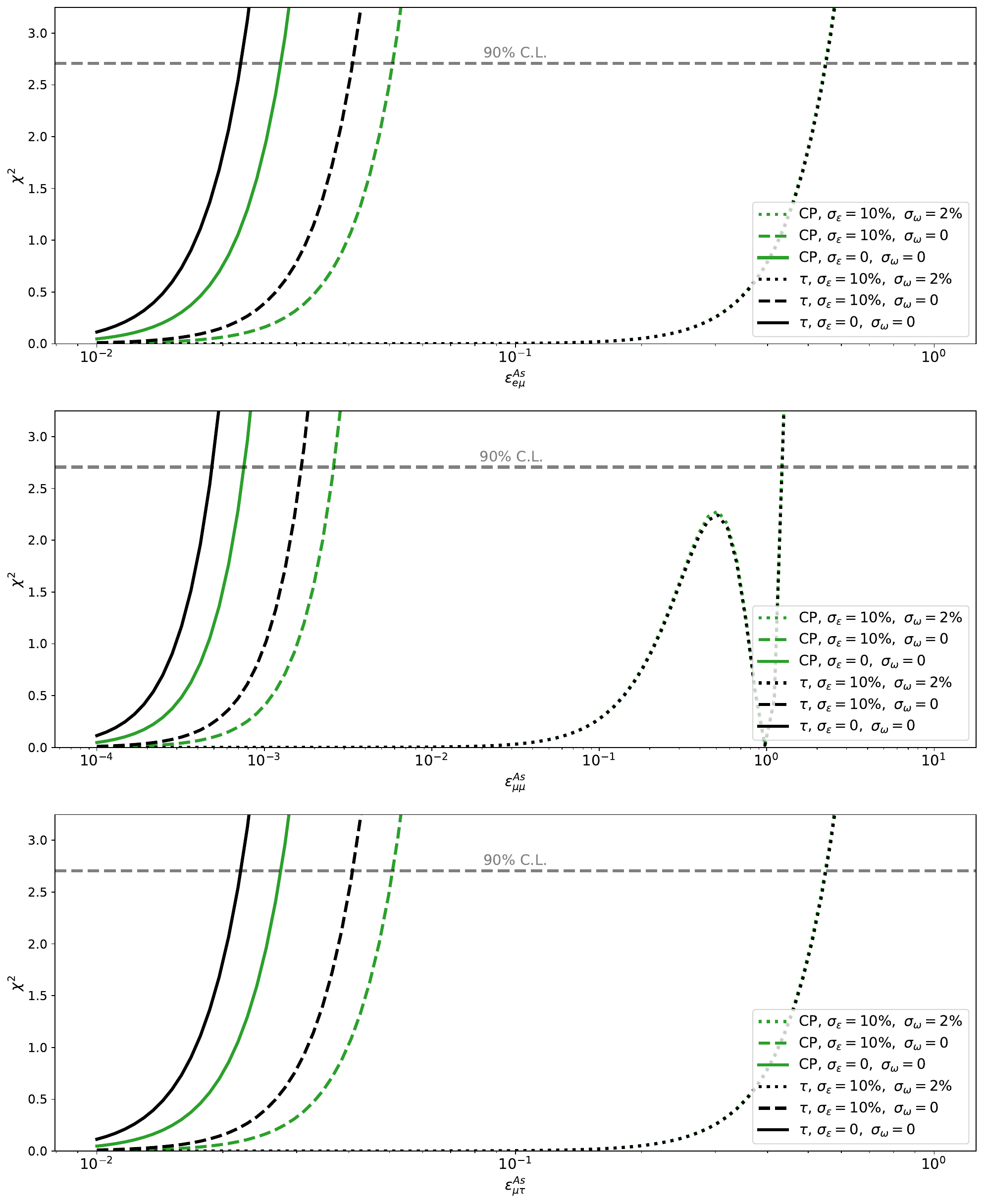}
    \caption{Similar to Fig.~\ref{fig:Chi2ND1u} for the neutral current axial NSI with the $s$ quark, $\epsilon^{As}$. The green (black) curves correspond to the CP-optimized ($\tau$-optimized) spectrum. The green dotted and the black dotted curves are similar.\label{fig:Chi2ND1s}}
\end{figure}
\begin{figure}
    \centering
    \includegraphics[height=0.82\textheight]{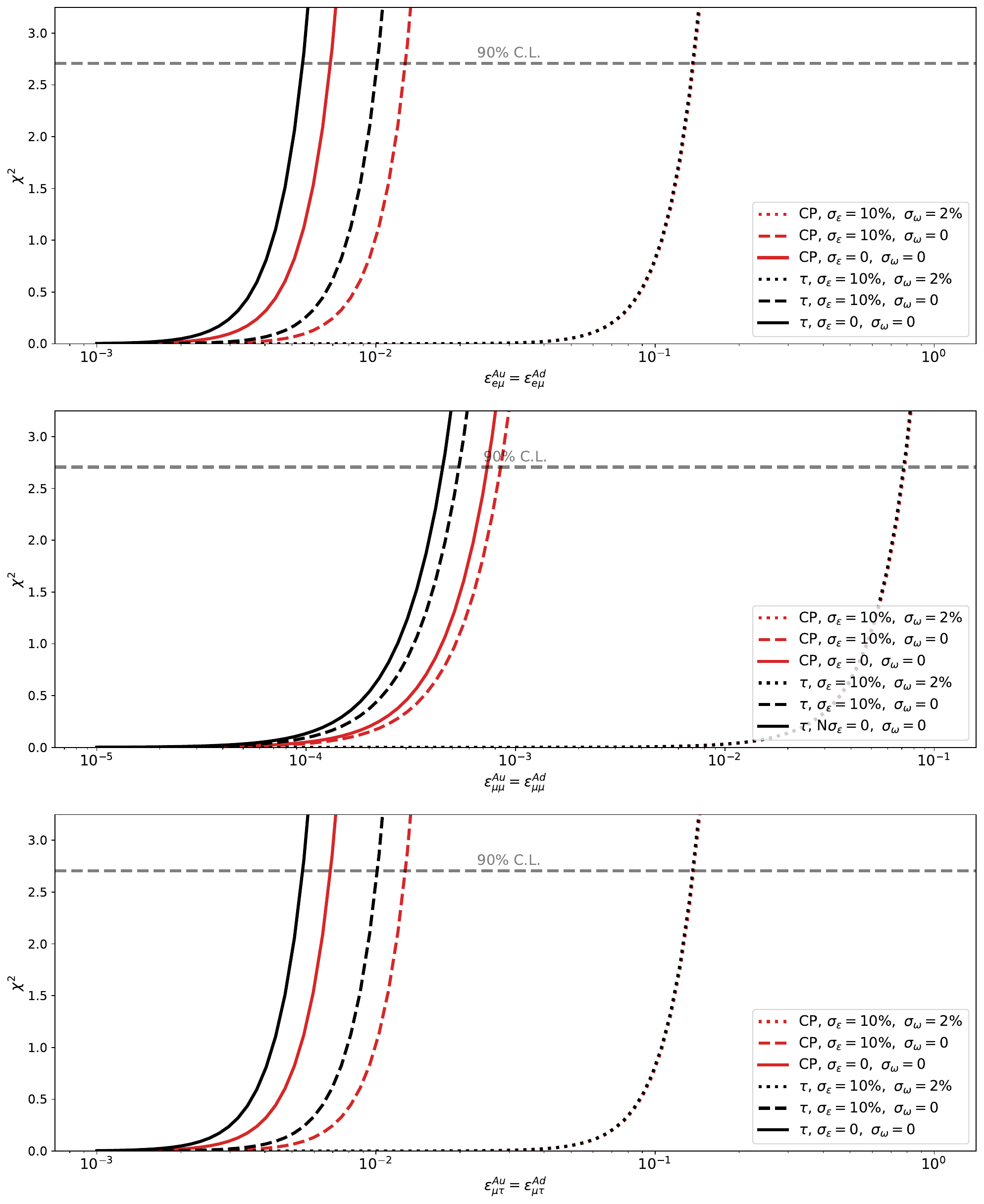}
    \caption{Similar to Fig.~\ref{fig:Chi2ND1u} except that the axial NSI with the $d$ and $u$ quarks are set to be equal to each other, $\epsilon^{Ad}=\epsilon^{Au}$. The red (black) curves correspond to the CP-optimized ($\tau$-optimized) spectrum. The red dotted and the black dotted curves are similar and indistinguishable. \label{fig:Chi2ND1ueqd}}
\end{figure}
\begin{figure}
    \centering
    \includegraphics[height=0.27\textheight]{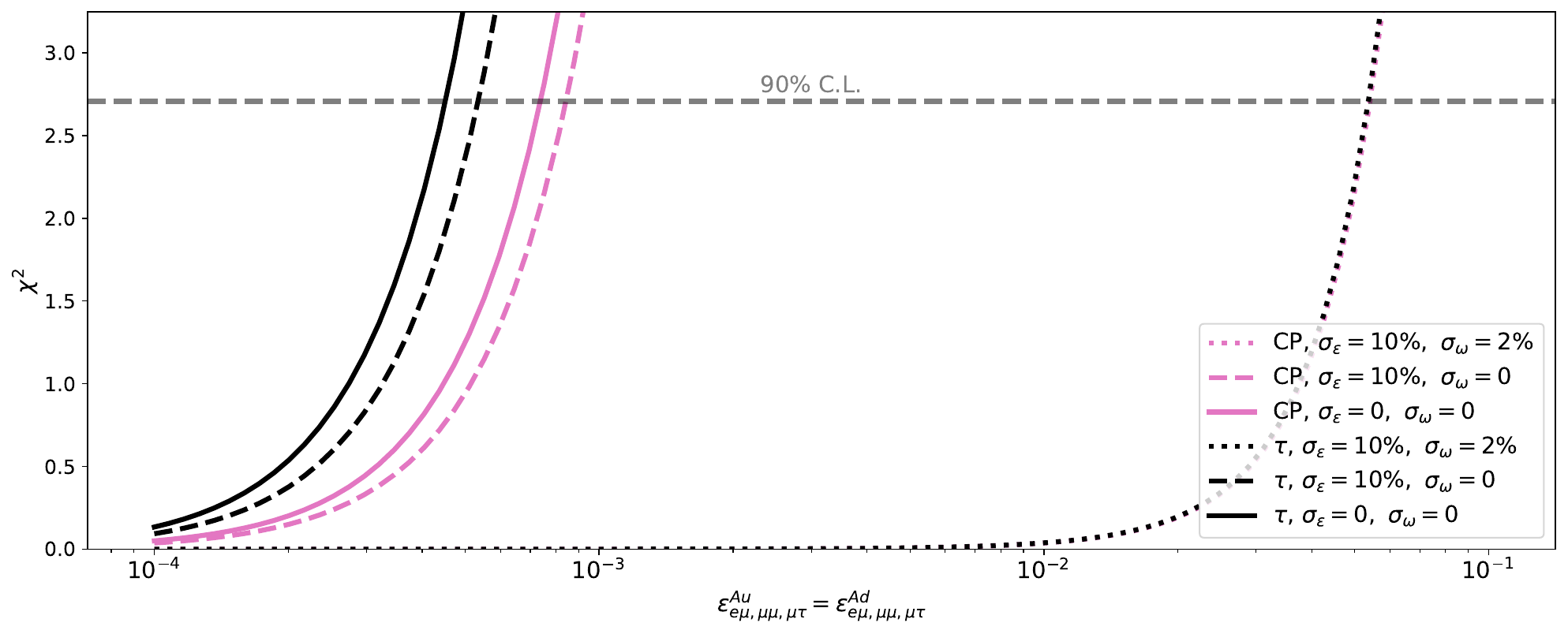}
    \caption{Similar to Fig.~\ref{fig:Chi2ND1ueqd} except that the $e \mu$ components of the axial NSI with the $d$ and $u$ quarks are set equal to the $\mu \mu$ and $\mu \tau$ components: $\epsilon^{Au}_{e\mu,\mu \mu,\mu\tau}=\epsilon^{Ad}_{e\mu,\mu \mu,\mu\tau}$. The pink (black) curves correspond to the CP-optimized ($\tau$-optimized) spectrum. The pink dotted and the black dotted curves are not distinguishable. \label{fig:Chi2ND3ueqd}}
\end{figure}

The 90\% bounds to be set by an experiment such as FD of DUNE after 6.5+6.5 years of data taking are summarized in table \ref{tab:bounds}. As seen from the figures as well as from the bounds in the table, the FD of a DUNE-like experiment can significantly improve the bounds on $\epsilon_{\tau\tau}^{Aq}$ and $\epsilon_{e\tau}^{Aq}$, even in the most pessimistic case of nonzero $\sigma_\omega$ and $\sigma_\epsilon$. The improvement is more significant for the case $\epsilon^{Au}=\epsilon^{Ad}$ where SNO is not able to set any bound.
Notice that the bound on $\epsilon_{\tau \tau}^{Aq}$ and $\epsilon_{e\tau}^{Aq}$ ($q\in \{ u,d \}$) can be significantly improved down to $\mathcal{O}(0.01)$ for $\sigma_\epsilon=0$ and down to $\mathcal{O}(0.1)$ for $\sigma_\epsilon=0.1$. The bound on $\epsilon_{ee}^{Aq}$ ($q\in \{ u,d \}$) can also be improved relative to the present bound from CHARM (see Eq.~(\ref{CHARM})) but as discussed in the previous section, the sensitivity of a DUNE-like experiment to the $ee$ component is limited by the smallness of $\sin \theta_{13}$. For the first time, we have considered the possibility to probe $\epsilon_{\alpha \beta}^{As}.$ It is impressive that a DUNE-like experiment can constrain its $\tau \tau$ component down to $\mathcal{O}(0.1)$ if $\sigma_\epsilon\ll 0.1$.

As mentioned before, in drawing Figs.~\ref{fig:Chi2FD1u}-\ref{fig:Chi2FD3ueqd}, the $\epsilon^{Aq}$ elements are set to be real. In general, $\epsilon_{\alpha \beta}^{Aq}$ ($\alpha\ne \beta$) can be complex. The scattering cross section of $\nu_\alpha +N \to \nu_\beta +X$ ($\alpha\ne \beta$, $\alpha,\beta\in \{e,\mu,\tau \}$) is given by $|\epsilon_{\alpha \beta}^{Aq}|^2$ and is independent of the phase of $\epsilon_{\alpha \beta}^{Aq}$. However, the scattering cross section of $\nu_{far} +N \to \nu_{e(\tau)}+X$ depends on the phase of  $\epsilon_{e \tau}^{Aq}$ because of the interference between the SM and NSI amplitudes when $\mathcal{A}_e$ and $\mathcal{A}_\tau$ are both nonzero. Fig. \ref{fig:Chi2FDContour1ueqd}  shows the 90 \% C.L. constraints in the $|\epsilon_{e \tau}^{Aq}|$
and ${\rm Arg}(\epsilon_{e \tau}^{Aq}$) plane by FD with the CP-optimized spectrum.

As discussed in sect. \ref{deutrerium}, there is a solution for SNO with large $\tau \tau$ component which is not still ruled out by any other experiment.
Fig. \ref{fig:r} shows how FD can test the $\epsilon_{\tau \tau}^{Au}-\epsilon_{\tau \tau}^{Ad}\simeq -1.5$ solution of SNO. The horizontal axis shows $r=\epsilon_{\tau \tau}^{Au}/\epsilon_{\tau \tau}^{Ad}$.
As $r\to 1$, the requirement $\epsilon_{\tau \tau}^{Au}-\epsilon_{\tau \tau}^{Ad}\simeq -1.5$ implies $\epsilon_{\tau \tau}^{Au},\epsilon_{\tau \tau}^{Ad}\to \infty$; as a result, at $r\to 1$, $\chi^2$ goes to infinity.
As seen in Fig. \ref{fig:r}, regardless of the value of $r$, the $\chi^2$ test can rule out the $\epsilon_{\tau \tau}^{Au}-\epsilon_{\tau \tau}^{Ad}\simeq -1.5$ solution of SNO with high confidence level.

In the above results, we have set the mixing parameters equal to their best fit according to the latest global neutrino data analysis. We checked the robustness of our results against varying $\delta $ in $(-\pi, \pi)$. As expected the change was negligible.

\begin{table}
	\caption{ Forecast for the 90\% C.L. bounds on axial NSI parameters at the FD by a DUNE-like experiment after 6.5 years of running in each neutrino and antineutrino modes. We have taken $ 1.1\times 10^{21}$ POT per year. For the nuisance parameter, we have taken ($\sigma_{\epsilon} = \sigma_{\omega} = 0$), ($\sigma_{\epsilon} = 10\%$, $\sigma_{\omega} = 0$), ($\sigma_{\epsilon} = 10\%$, $\sigma_{\omega} = 2\%$) at the third, fourth, and fifth columns, respectively. All the elements are taken to be real.
		\label{tab:bounds}}
	\begin{ruledtabular}
\begin{tabular}{ c c c c c } 
	Parameter&Flux& $\sigma_{\epsilon}=0$, $\sigma_{\omega}=0$ & $\sigma_{\epsilon}=10\%$, $\sigma_{\omega}=0$  & \qquad$\sigma_{\epsilon}=10\%$, $\sigma_{\omega}=2\%$ \\
	\hline
	\multirow{2}{2em}{$\epsilon_{ee}^{Au}$} & CP & [-1.28, 0.23] & [-2.7, 0.9]& [-2.8, 1.2]\\ 
	& $\tau$ & [-1.35, 0.33] & [-2.1, 2.5] & [-2.9, 2.7] \\ 
	\hline
	\multirow{2}{2em}{$\epsilon_{e\tau}^{Au}$} & CP & [-0.065, 0.038] & [-0.22, 0.23]& [-0.26, 0.29]\\ 
	& $\tau$ & [-0.078, 0.072]& [-0.22, 0.30] & [-0.32, 0.40] \\ 
	\hline
	\multirow{2}{2em}{$\epsilon_{\tau\tau}^{Au}$} & CP & [-0.014, 0.014] & [-0.082, 0.072]& [-0.118, 0.100]\\ 
	& $\tau$ & [-0.021, 0.021] & [-0.652,0.503]+[-0.117, 0.089] & [-0.718, 0.156] \\ 
	\hline\hline
	\multirow{2}{2em}{$\epsilon_{ee}^{Ad}$} & CP & [-0.20, 0.35]$+$[0.93, 1.24] & [-0.7, 2.7]& [-0.96, 2.78]\\ 
	& $\tau$ & [-0.30, 1.29] & [-2.8, 1.2] & [-2.77, 2.37] \\ 
	\hline
	\multirow{2}{2em}{$\epsilon_{e\tau}^{Ad}$} & CP & [-0.051,0.040] & [-0.18,  0.23]& [-0.23, 0.27]\\ 
	& $\tau$ &[-0.076, 0.052] & [-0.24, 0.23] & [-0.33, 0.32] \\ 
	\hline
	\multirow{2}{2em}{$\epsilon_{\tau\tau}^{Ad}$} & CP & [-0.014, 0.016] & [-0.11, 0.24]& [-0.14, 0.24]\\ 
	& $\tau$ & [-0.021, 0.021] & [-0.14, 0.37] & [-0.21, 0.44] \\ 
	\hline\hline
		\multirow{2}{2em}{$\epsilon_{ee}^{As}$} & CP & [-1.1, 2.1] & [-4.1, 5.0]& [-4.7, 5.7]\\ 
	& $\tau$ & [-1.5, 2.5] & [-6.1, 7.1] & [-7.6, 8.6] \\ 
	\hline
	\multirow{2}{2em}{$\epsilon_{e\tau}^{As}$} & CP & [-0.54, 0.22] & [-0.72, 0.52] & [-0.85, 0.65]\\ 
	& $\tau$ & [-0.59, 0.27] & [-0.82, 0.57] & [-1.1,  0.84] \\ 
	\hline
	\multirow{2}{2em}{$\epsilon_{\tau\tau}^{As}$} & CP & [-0.11, 0.15]$+$ [0.85, 1.11] & [-0.29, 1.28]& [-0.39, 1.39]\\ 
	& $\tau$ & [-0.14, 0.21]$+$[0.79, 1.45] & [0.35, 1.35] & [-0.58, 1.58] \\ 
	\hline\hline
	\multirow{2}{4em}{$\epsilon_{ee}^{Au}$=$\epsilon_{ee}^{Ad}$} & CP & [-0.35, 0.41]& [-0.9, 1.3]& [-1.1, 1.5]\\ 
	& $\tau$ & [-0.45, 0.49] & [-1.9, 1.3] & [-2.1, 1.8] \\ 
	\hline
	\multirow{2}{4em}{$\epsilon_{e\tau}^{Au}$=$\epsilon_{e\tau}^{Ad}$} & CP & [-0.089, 0.083] & [-0.13, 0.18]& [-0.16, 0.21]\\ 
	& $\tau$ & [-0.101, 0.095] & [-0.14, 0.20] & [-0.21, 0.26] \\ 
	\hline
	\multirow{2}{4em}{$\epsilon_{\tau\tau}^{Au}$=$\epsilon_{\tau\tau}^{Ad}$} & CP & [-0.076, 0.084] & [-0.24, 0.09]& [-0.27, 0.13]\\ 
	& $\tau$ & [-0.088, 0.099] & [-0.26, 0.11] & [-0.32, 0.18] \\ 
\end{tabular}
\end{ruledtabular}
\end{table}

\begin{table}
	\caption{ Forecast for the 90\% C.L. bounds on axial NSI parameters at the ND by a DUNE-like experiment after 6.5 years of running in each neutrino and antineutrino modes. We have taken $ 1.1\times 10^{21}$ POT per year and an on-axis near detector of 67.2 ton of LAr. For the nuisance parameter, we have taken ($\sigma_{\epsilon} = \sigma_{\omega} = 0$), ($\sigma_{\epsilon} = 10\%$, $\sigma_{\omega} = 0$), ($\sigma_{\epsilon} = 10\%$, $\sigma_{\omega} = 2\%$) at the third, fourth, and fifth columns, respectively. All the elements are taken to be real.
		\label{tab:bounds2}}
	\begin{ruledtabular}
		\begin{tabular}{ c c c c c } 
			Parameter&Flux& $\sigma_{\epsilon}=0$, $\sigma_{\omega}=0$ & $\sigma_{\epsilon}=10\%$, $\sigma_{\omega}=0$  & \qquad$\sigma_{\epsilon}=10\%$, $\sigma_{\omega}=2\%$ \\
			\hline
			\multirow{2}{2em}{$\epsilon_{e\mu}^{Au}$} & CP & [-0.0098, 0.0098] & [-0.018,0.018]& [-0.19,0.19]\\ 
			& $\tau$ & [-0.0078,0.0078] & [-0.015,0.015] & [-0.19,0.19] \\ 
			\hline
			\multirow{2}{2em}{$\epsilon_{\mu\mu}^{Au}$} & CP & [-0.000099,0.000099] & [-0.0007,0.0007]& [-0.065,0.065]\\ 
			& $\tau$ & [-0.000062,0.000062] & [-0.0004,0.0004] & [-0.065,0.065] \\ 
			\hline
			\multirow{2}{2em}{$\epsilon_{\mu\tau}^{Au}$} & CP & [-0.0098,0.0098] & [-0.018,0.018]& [-0.19,0.19]\\ 
			& $\tau$ & [-0.0078,0.0078] & [-0.014,0.014] & [-0.19,0.19] \\ 
			\hline\hline
			\multirow{2}{2em}{$\epsilon_{e\mu}^{Ad}$} & CP & [-0.0096,0.0096] & [-0.018,0.018]& [-0.19,0.19]\\ 
			& $\tau$ & [-0.0076,0.0076] & [-0.014,0.014] & [-0.19,0.19] \\ 
			\hline
			\multirow{2}{2em}{$\epsilon_{\mu\mu}^{Ad}$} & CP & [-0.00009,0.00009] & [-0.0034,0.0034]& [-0.12,0.12]\\ 
			& $\tau$ & [-0.00006,0.00006] & [-0.0021,0.0021] & [-0.12,0.12] \\ 
			\hline
			\multirow{2}{2em}{$\epsilon_{\mu\tau}^{Ad}$} & CP & [-0.0095,0.0095] & [-0.018,0.018]& [-0.19,0.19]\\ 
			& $\tau$ & [-0.0076,0.0076] & [-0.014,0.014] & [-0.19,0.19] \\ 
			\hline\hline
			\multirow{2}{2em}{$\epsilon_{e\mu}^{As}$} & CP & [-0.027,0.027] & [-0.051,0.051]& [-0.55,0.55]\\ 
			& $\tau$ & [-0.022,0.022] & [-0.041,0.041] & [-0.55,0.55]\\ 
			\hline
			\multirow{2}{2em}{$\epsilon_{\mu\mu}^{As}$} & CP & [-0.00075,0.00075] & [-0.0026,0.0026]& [-1.21,1.21]\\ 
			& $\tau$ & [-0.00048,0.00048] & [-0.0016,0.0016] & [-1.21,1.21] \\ 
			\hline
			\multirow{2}{2em}{$\epsilon_{\mu\tau}^{As}$} & CP & [-0.027,0.027] & [-0.051,0.051]& [-0.55,0.55]\\ 
			& $\tau$ & [-0.022,0.022] & [-0.041,0.041] & [-0.55,0.55] \\ 
			\hline\hline
			\multirow{2}{4em}{$\epsilon_{e\mu}^{Au}$=$\epsilon_{e\mu}^{Ad}$} & CP & [-0.0069,0.0069] & [-0.013,0.013]& [-0.14,0.14]\\ 
			& $\tau$ & [-0.0055,0.0055] & [-0.010,0.010] & [-0.14,0.14] \\ 
			\hline
			\multirow{2}{4em}{$\epsilon_{\mu\mu}^{Au}$=$\epsilon_{\mu\mu}^{Ad}$} & CP & [-0.00074,0.00074] & [-0.00085,0.00085]& [-0.072,0.072]\\ 
			& $\tau$ & [-0.00045,0.00045] & [-0.00054,0.00054] & [-0.072,0.072] \\ 
			\hline
			\multirow{2}{4em}{$\epsilon_{\mu\tau}^{Au}$=$\epsilon_{\mu\tau}^{Ad}$} & CP & [-0.0069,0.0069] & [-0.013,0.013]& [-0.14,0.14]\\ 
			& $\tau$ & [-0.0055,0.0055]& [-0.010,0.010] & [-0.14,0.14] \\ 
		\end{tabular}
	\end{ruledtabular}
\end{table}

Let us now discuss the possibility of constraining $\epsilon^{Aq}_{\mu \alpha}$. Having a huge flux of $\nu_\mu$ and $\overline{\nu}_\mu$ at ND ($\mathcal{N}^{\rm ND}\sim \mathcal{O}(10^8)$), there is an excellent prospect of improving the bounds on these elements. We again define $\chi^2$ as in Eqs. (\ref{chiNOT}) replacing FD$\to$ND. 
The results are shown in Figs.~\ref{fig:Chi2ND1u}-\ref{fig:Chi2ND3ueqd}.
As seen from the figures in table \ref{tab:bounds2}, in the absence of systematic errors ({\it i.e.,} $\sigma_{\epsilon}=0$ and $\sigma_{\omega}=0$), $\epsilon_{\mu \mu}^{Au/d}$ can be constrained down to $\mathcal{O}(10^{-4})$ but $\epsilon_{e \mu}^{Au/d}$ and $\epsilon_{\tau \mu}^{Au/d}$ can be constrained only down to $\mathcal{O}(10^{-2})$. The reason for relatively weaker bounds on the lepton flavor violating components is that while for the $\mu \mu$ component, there is an interference between NSI and standard amplitudes, giving rise to a linear term in $\epsilon_{\mu\mu}^{Au/d}$ in the cross section, there is no such interference for the lepton flavor violating components and the leading contributions to $\epsilon_{\mu e}^{Au/d}$ and/or $\epsilon_{\mu \tau}^{Au/d}$ are only quadratic.

As indicated in table \ref{tab:bounds2},turning on the systematic error $\sigma_\epsilon$, the bounds become  weaker but with $\sigma_\omega=0$ (or with correlated $\omega=\omega_\nu=\omega_{\bar{\nu}}$ with $\sigma_\omega=0.02$) $\epsilon_{\mu \mu}^{A u/d}$ can still be constrained down to $\mathcal{O}(10^{-3})$. Notice that we have assigned a single nuisance parameter to the neutrino and antineutrino modes for the treatment of the uncertainty of $\epsilon$ with the justification that the dominant errors such as miscalculation of nuclear effects or the errors in the evaluation of the signal efficiency will be equal for the both modes. If we assigned two separate pull parameters for the neutrino and antineutrino with $\sigma_\epsilon^\nu=\sigma_\epsilon^{\overline{\nu}}=\mathcal{O}(0.1)$, the bound on the $\mu\mu$ component could not be better than $\mathcal{O}(0.1)$. Moreover,  taking uncorrelated  uncertainties
in the  backgrounds for the neutrino and antineutrino modes ({\it i.e.,} introducing two independent nuisance parameters 
 $\omega_{\nu}$ and $\omega_{\bar{\nu}}$) deteriorates the forecast for the bounds.  We also examined the case where the backgrounds of neutrinos and antineutrinos are correlated so their uncertainty can be treated with a single nuisance parameter, $\omega_{\nu} = \omega_{\bar{\nu}}$ and found that the forecasted bounds at both ND and FD will not be much different from the case that only the uncertainty of efficiency is non-zero.
Notice that the bounds with the $\tau$-optimized spectrum (shown by the black curves) are expected to be stronger than those with the CP-optimized spectrum (shown with the colored curves). This is understandable as with the harder spectrum of the $\tau$-optimized mode, the cross section and therefore the statistics will be higher. The forecast bounds are summarized in tables \ref{tab:bounds} and \ref{tab:bounds2}. Notice that the bounds from a DUNE-like experiment will be far stronger than those from FASER$\nu$ \cite{Escrihuela:2023sfb}. 

Since the flux at the near detector is a pure flavor state, the $\nu_\mu +N \to \nu_{e(\tau)}+X$ amplitude does not receive a contribution from the SM. As a result, there is no interference in the cross section of $\nu_\mu +N \to \nu_{e(\tau)}+X$ and the relevant cross section is proportional to $|\epsilon^{Aq}_{\mu e(\tau)}|^2$ and is therefore independent of Arg[$\epsilon^{Aq}_{\mu e(\tau)}$].

\section{Summary}
\label{sec:summary}
We have studied how a DUNE-like experiment can determine the axial NC NSI parameters. Taking a general lepton flavor structure for NSI, we have first derived the cross sections of NC DIS in the presence of NSI for a definite flavor of the incoming neutrinos and have then discussed cross section of an oscillated flux composed of coherent combination of all three flavors. We have also shortly discussed how axial NC NSI affects the resonant neutrino nucleon scattering. We have studied the variation of the number of NC DIS events with $\epsilon_{\alpha \beta}^{Aq}$. The events at the near detector is sensitive only to $\epsilon_{\alpha \mu}^{Aq}$ because the flux reaching near detector is mainly composed of $\nu_\mu$ or $\overline{\nu}_\mu$. However, the far detector can be sensitive to all flavor elements of $\epsilon_{\alpha \beta}^{Aq}$ because the flux reaching the FD is composed of all flavors due to oscillation. However, as expected the sensitivity to $\epsilon_{ee}^{Aq}$ is lower than other elements simply because $P(\nu_\mu \to \nu_e)$ is smaller than $P(\nu_\mu \to \nu_\tau)$. Again as expected, the sensitivity to $\epsilon^{As}_{\alpha \beta}$ is lower than that to $\epsilon^{Au}_{\alpha \beta}$ or $\epsilon^{Ad}_{\alpha \beta}$ because there is no valence $s$-quark in the nucleons.

We have made a forecast for bounds that can be set on $\epsilon_{\alpha \beta}^{Aq}$ for different ratios of $\epsilon^{Au}/\epsilon^{Ad}$ with a DUNE-like setup after 6.5+6.5 years of running in the neutrino and antineutrino modes. The near detector which enjoys detection of $\mathcal{O}(10^8)$ NC DIS events will have enough statistics to probe $\epsilon^{A u/d}_{\mu\mu}$ down to $\mathcal{O}(10^{-4})$ and $\epsilon^{A u/d}_{\mu\alpha}$ ($\alpha \in \{ e ,\tau \}$) down to $\mathcal{O}(10^{-2}).$ However, the systematic errors limit the sensitivity. 
We have shown that if the systematic errors in the neutrino and antineutrino modes are the same ({\it i.e.,} if they can be treated with a single pull parameter), the sensitivity to $\epsilon_{\mu \mu}^{A u/d}$ (to $\epsilon^{A u/d}_{\mu\alpha}$, $\alpha \in \{ e ,\tau \}$) still remains below $\sim 5\times 10^{-3}$ ($\sim 5\times 10^{-2}$) even if the systematic errors in the flux prediction is $\mathcal{O}(10\%)$. We found that the bounds will not significantly change even in the presence of correlated errors in the background estimation for the neutrino and antineutrino modes. However, in the pessimistic and unlikely case that the background estimation errors for the neutrino and antineutrino modes are not correlated and are therefore treated with two independent pull parameters, the bounds become significantly deteriorated.

The far detector can constrain $\epsilon_{\tau \tau}^{Aq}$, $\epsilon_{e \tau}^{Aq}$ and $\epsilon_{ee}^{Aq}$. The bounds for different ratios of $\epsilon^{Au}/\epsilon^{Ad}$ are shown in table \ref{tab:bounds}. Of particular interest is the case $\epsilon_{\tau \tau}^{Au}=\epsilon_{\tau \tau}^{Ad}$ which is by now practically unconstrained. As seen in the table, a DUNE-like experiment can constrain the $\tau \tau$ component to $\mathcal{O}(0.1)$ (below $\mathcal{O}(0.1)$) with (without) a systematic error of 10 \% in the signal prediction.
We have also shown that the non-trivial (still allowed) SNO solution with $\epsilon^{Au}_{\tau\tau}-\epsilon^{Ad}_{\tau\tau}=-1.5$ can be ruled out by a DUNE-like experiment with high confidence level for any $\epsilon^{Au}/\epsilon^{Ad}$ value.
For the first time, we have examined the possibility to probe the axial NSI of neutrinos with the $s$-quark. As seen in table \ref{tab:bounds}, all $\epsilon_{\alpha \beta}^{As}$ components except $\epsilon_{ee}^{As}$ can be strongly constrained by a DUNE-like experiment.

We have carried out our analysis for both CP-optimized and $\tau$-optimized spectra. As expected, with equal POT per year for both cases, the sensitivity of near detector to $\epsilon^{Aq}_{\mu \alpha}$ will be higher with the $\tau$-optimized spectrum which has a higher average energy and therefore yields a higher number of events. However, the sensitivity of the far detector to $\epsilon_{\tau \tau}^{Aq}$ deteriorates for the $\tau$-optimized spectrum because $P(\nu_\mu \to \nu_\tau)$ is reduced at high energies relevant for the $\tau$-optimized spectrum.

\acknowledgments
The authors are grateful to Hamed Abdolmaleki, Saeed Ansarifard, Hadi Hashamipour, and Michel Sorel for the valuable discussions. SS and MD would like to thank IPM for partial financial support through IPM grant of Y. Farzan.
This work has been supported by the European Union's Framework Programme for Research and Innovation Horizon 2020 under the Marie Skłodowska-Curie grant agreement No 860881-HIDDeN as well as under the Marie Sklodowska-Curie Staff Exchange grant agreement No 101086085-ASYMMETRY.
YF would like to acknowledge support from ICTP through the Associates Programme and from the Simons Foundation through grant number 284558FY19. This work represents the views of the authors and should not be considered a DUNE collaboration paper.

\section*{Code Availability Statement}
Our codebase computing the number of Neutral Current Deep Inelastic Scattering events in the presence of Axial Non-Standard Interaction at near and far detectors of a DUNE-like setup is publicly available at \href{https://github.com/dehpour/AxialNCNSI-DUNE/}{https://github.com/dehpour/AxialNCNSI-DUNE/}.

\bibliography{biblio}

\end{document}